%% file: main.tex
\documentclass{tlp}

\def \filename{main}
\usepackage{Summary_Styles}
\usepackage{Summary_Commands}

\begin{document}

\input frontmatter.tex

\input intro.tex
\input wspl.tex

\input contexts.tex

\input datareft.tex

\input modreft.tex

\input deriv.tex

\input demonic-deriv.tex

\input relwork.tex

\input concl.tex

\bibliographystyle{acmtrans}
\bibliography{biblio,colvinpubs}

\end{document}

%% file: frontmatter.tex
\maketitle

\centerline{\it \small To appear in Theory and Practice of Logic Programming
(TPLP)}

\begin{abstract}

The refinement calculus for logic programs is a framework for deriving logic
programs from specifications.  It is based on a wide-spectrum language that
can express both specifications and code, and a refinement relation that
models the notion of correct implementation.  
In this paper we extend and generalise earlier work on \emph{contextual
refinement}. 
Contextual refinement 
simplifies the refinement process by abstractly
capturing the context of a subcomponent of a program,
which typically includes information about the values of the free variables.
This paper also extends and generalises \emph{module refinement}.
A \emph{module} is a collection of procedures that operate 
on a common data type;
module refinement between a specification
module $A$ and an implementation module $C$
allows calls to the procedures of $A$ to be systematically replaced with
calls to the corresponding procedures of $C$.
Based on the conditions for module refinement, we present a method for
\emph{calculating} an implementation module from a specification module.
Both contextual and
module refinement within the refinement calculus have been generalised from
earlier work and the results are presented in a unified framework.

\COMMENT{
Because our wide-spectrum language is monotonic with respect to the
refinement ordering, 
a program, $S$, is refined by refining any of its components.
We can use this property to decompose the refinement of a program into the
refinement of (some or all of) its components.
In many situations a component of $S$, $c$, may inherit context from
$S$.  
This context can, for example,
provide information about the values of free variables in the
component.
In this paper we provide a framework for
making context available during the refinement of 
a program's components.

We use contextual refinement to reason about module refinement.  A
module in our language is a group of procedures that operate on a common
data type.  By
making assumptions about the structure of a program that uses the module we
derive a context in which efficient implementations of abstract data
types are allowed.  Finally, we present a method for deriving, or
\emph{calculating}, an implementation module from an abstract module.
Starting from the abstract module and a relation between the abstract and
implementation types that satisfies some consistency checks, a
specification of the implementation module can be automatically produced.
}


\end{abstract}

%% file: intro.tex
\section{Introduction}

The construction of
programs that are correct with respect to their specifications is an
important goal of software development.
A \emph{refinement calculus} is a formal method for deriving programs from
specifications in a step-wise fashion.
It is based on:
\begin{itemize}
\item
   a {\em wide-spectrum language\/}
    that can express both specifications and executable programs;
\item
   a {\em refinement relation\/} that
    models the notion of correct implementation; and
\item
   a collection of {\em refinement laws\/} providing the means to
    refine specifications to code in a stepwise fashion.
\end{itemize}

The wide-spectrum language contains both specification and implementation
constructs, blurring the distinction between
specifications and executable code.
A series of correctness-preserving refinement laws are applied to a
specification, replacing specification constructs with implementation
constructs.
Each refinement law is
proved with respect to the underlying semantics of the calculus.
A law may have associated \emph{proof obligations}, which must be
discharged to ensure the application of the law is valid.

\COMMENT{
A specification and its implementation as a logic program are, in
general, more closely related than a comparable imperative
specification and its implementation,
since in the logic programming paradigm both specifications and implementations
may be defined in terms of the same
logical notation.
This reduced gap between a specification and its
implementation should result in a shorter refinement process.
}

A refinement calculus for logic programs has been developed
\cite{reflp:96,semanticsTR:00,reflp:01}.  
In this paper we extend and generalise earlier work on \emph{contextual}
and \emph{module} refinement of logic programs within the refinement
calculus,
and present the results in a unified framework.

Because our wide-spectrum language is monotonic with respect to the
refinement ordering,
a program, $S$, is refined by refining any of its components.
We can use this property to decompose the refinement of a program into the
refinement of (some or all of) its components.
In many situations a component of $S$ may inherit context from
$S$.
This context can, for example,
provide information about the values of free variables in the
component.
In this paper we provide a framework for
making context available during the refinement of
a program's components.

We use contextual refinement to reason about module refinement.  A
module in our language is a group of procedures that operate on a common
data type.  By
making assumptions about the structure of a program that uses the module we
derive a context in which efficient implementations of abstract data
types are allowed.  Finally, we present a method for deriving, or
\emph{calculating}, an implementation module from an abstract module.
Starting from the abstract module and a \emph{coupling invariant}
--- a relation between the abstract and
implementation types --- a
specification of the implementation module can be automatically produced
(subject to some consistency checks).


\COMMENT{
A refinement calculus for logic programs has been developed
\cite{reflp:96,semanticsTR:00,reflp:01}.
In this paper, we focus on contextual and module refinement.
The notion of data refinement, where the type of a variable is changed, has
been explored in \cite{datareft:fmp98:full} and \cite{modreft:lopstr00}.
The treatment of types, and their generalisation, invariants, in terms of
providing
context for the refinement of programs has been discussed in
\cite{typesinv:acsc00}.
This paper combines the above results and summarises the contributions of
\cite{Thesis}.
A consistent structure and notation has been adopted for each topic,
resulting in a simpler and more comprehensive theory for contextual and module
refinement.
Specifically,
it generalises the results of \cite{typesinv:acsc00} by unifying
the treatment of context for the different constructs in the language.
The results of \cite{datareft:fmp98:full} are condensed and simplified in the
unified notation.
It expands on the results of \cite{modreft:lopstr00} by considering more
program structures and allowing arbitrary predicates as context.
A more involved example is used to present the results.
Finally, a novel technique for automatically
calculating implementation modules is presented.
}

The paper is structured as follows.
In \refsect{wspl} the meaning of wide-spectrum language constructs
and refinement are informally described.
\refsect{context} examines contextual refinement of logic
programs.  
The contextual refinement laws
are illustrated with an example of a data refinement.
In \refsect{modreft} we discuss module refinement, where 
we reason about 
groups of procedures that operate on a common data type.
In \refsect{modderiv} we present a general scheme for deriving an
implementation of a module based on the relationship between the
specification and implementation types.
We then specialise the scheme for particular combinations of
abstract operations and coupling invariants.
In particular,
\refsect{demonic-deriv} extends the specification 
language so that nondeterminism in some coupling invariants
can be eliminated,
allowing more efficient implementation modules.
In \refsect{relwork} we discuss related work.

This paper summarises and extends
the first author's thesis \cite{Thesis}.
We combine and extend the results of earlier papers 
\cite{datareft:fmp98:full,typesinv:acsc00,modreft:lopstr00} and adopt a
consistent structure and notation, resulting in a simpler and more
comprehensive theory for contextual and data refinement.
Specifically,
the results of \citeN{typesinv:acsc00} are generalised by unifying the
treatment of context for the different constructs in the language
(\refsect{context}), and the
results of \citeN{datareft:fmp98:full} are condensed and simplified in the
unified notation (\refsect{reverse:dr}).  
The results of \citeN{modreft:lopstr00} are extended by
considering more program structures and allowing arbitrary predicates as
context, and a more complex example is used to present the results
(\refsect{modreft}).
We also present a
technique for automatically calculating implementation modules (\refsect{modderiv}), originally
proposed in \citeN{Thesis}. 
Specialisations of the calculation technique (\refsect{sect-specialisations})
and the use of demonic nondeterminism in module calculations
(\refsect{demonic-deriv}) are novel to this paper.

\COMMENT{
For brevity of presentation, 
in this paper we express the important theorems in terms of refinement laws,
rather than directly in the underlying semantics.
All laws used have been proved correct with respect to the semantics
described in \citeN{reflp:01}.
}

\COMMENT{
In this paper we avoid the presentation of formal semantics for the language,
and instead express the important theorems in terms of refinement laws.
These refinement laws have been proved correct with respect to the semantics
presented in \cite{reflp:01}.
}



\COMMENT{
The paper is concluded in \refsect{conclusions} and future work is
discussed.  \refappendix{ax} contains refinement and predicate
laws used in the paper (proofs for all laws used may be found in
\cite{Thesis} or \cite{TPLP}).
Laws introduced in a section are repeated in the
appendix; any use of a law from outside the section refers to the appendix
version.
\refappendix{module:reft:proof} 
presents...
\refappendix{setlist:ax} 
presents...
}

\COMMENT{
In \refsect{relwork} we compare the refinement calculus approach to that of
deductive synthesis.
Work related to a specific topic of the paper
is mentioned at the end of the appropriate section.
}

\COMMENT{
The paper is organised as follows.
\refsect{wspl} introduces our wide-spectrum language.
\refsect{semantics} gives a semantics for the language and refinement.
\refsect{context} extends the semantics to introduce simplifications
when using context in refinements.
\refsect{datareft} introduces the notion of data refinement of
individual procedures.
\refsect{modreft} introduces the notion of data refinement of groups
of procedures (modules), provided that calls to the module obey a
particular structural form.
In \refsect{modderiv} we provide a general scheme for deriving an
implementation of an abstract module.  We specialise this scheme based
on the relationship between the specification and implementation
types.
}

\COMMENT{
In particular, we examine the difference between programs conjoined
sequentially
and programs conjoined in parallel.  In the former case, the second
program may assume that the first has successfully executed, and the context
it has established.
In the latter, the context is more complicated, though
parallel conjunction has the advantage of being more abstract than sequential
conjunction as a program operator.
%
This section extends and
generalises earlier work \cite{tool:nfmw97, typesinv:acsc00}, by considering
more program structures and considering arbitrary predicates as context rather
than limiting the discussion to types.
}
\COMMENT{
\refsect{datareft}
looks at \emph{data refinement}, which is a method for
replacing a data type within a program with another data type,
maintaining the original meaning of the program.
The original data type, referred to as the \emph{abstract} type, is
related to the new, or \emph{concrete}, data type via a \emph{coupling
invariant}.
The section describes three cases for data refinement on individual
procedures,
distinguished by interface issues.
}
\COMMENT{
The first case is a data refinement in context;
a call on an abstract procedure may be refined to a call on a concrete
procedure provided that the coupling invariant is established in the
context of the call.
The signature of the abstract and concrete procedures is different,
since they expect abstract and concrete types respectively as
parameters.
The second case of data refinement is similar, except that the
abstract signature must be maintained;
thus, the data refinement is hidden from the calling program.  In the third
case, a variable local to the procedure is data refined.
An earlier version of this work appears in \cite{datareft:fmp98:full}.
}

%% file: wspl.tex
\section{The wide-spectrum language and refinement}
\label{wspl} 

A wide-spectrum language 
may be used to express both specifications as well as
executable programs \cite{Partsch-90}.  
For example, \citeN{Back:88} included
specification constructs in Dijkstra's imperative language
\cite{Dijkstra:76}.
Using a wide-spectrum language
has the benefit of allowing stepwise refinement within a single
notational framework.

\subsection{Basic constructs}

\paragraph{Semantic model.}
For brevity we present an informal, intuitive description of the
semantics of the language and refinement, and
present the main theorems and results as high-level refinement laws. 
The details of
a predicate-based semantics appears in \cite{reflp:96}, and
of an operational semantics in \cite{reflp:01}.

In our language, a $command$ (logic program fragment) $S$ 
with free variables $V$
constrains (instantiates) $V$ to satisfy $S$.
(This is the same principal involved as when a procedure call $p(V)$
constrains $V$ to satisfy $p$.)
The instantiation of the free variables, which may already be partially or
fully instantiated, is the ``effect'' of $S$, similar to a postcondition in
Hoare logic.
Additionally,
every command may have an associated ``assumption'', similar to preconditions in
Hoare logic.
Assumptions
specify the
instantiations of the free variables for which the command is guaranteed
to function correctly.
If the free variables do not satisfy the assumptions,
the program may behave in any manner (like $\Abort$ in
Dijkstra's language).

The commands in our wide-spectrum language are discussed below
(a summary appears in Fig.~\ref{wide-spec-lang}).
We describe them in terms of their assumptions (input instantiations) and
effect (output instantiations).
Throughout the paper we adopt the following naming conventions.
\[
\begin{array}{ll}
	A,B.. & \mbox{predicates (inside \emph{assumption} commands)} \\
	P,Q.. & \mbox{predicates (inside \emph{specification} commands)} \\
	S,T.. & \mbox{commands} \\
	V,X,Y & \mbox{variables} \\
	U & \mbox{terms}
\end{array}
\]

\COMMENT{
, $p$, when presented with some free variables (which are
possibly partially instantiated), returns all instantiations of the free
variables which are consistent with the initial instantiations and also
satisfy $p$.  Hence we can informally think of logic programs as partial
functions from instantiations to instantiations (where the set of output
instantiations is a subset of the set of input instantiations).
The functions are partial because a program may be undefined for some
inputs, for example the division program may be undefined if the divisor
parameter is 0.
We describe the commands in our language with respect to this informal
treatment of logic programs.
}

\COMMENT{
A query \T{q(X)} in a logic programming language such as Prolog
can be understood as finding all instantiations of $X$ that satisfy the
program \T{q}.  Hence, we model programs as functions that manipulate
\emph{States}, which are mappings from variables to sets of values (the
values that satisfy the program).
The input to these functions (programs) are states, which may already
provide some bindings for the free variables.
In the query \T{q(X)}, \T{X} is not bound, and hence is initially mapped to
all possible values.  A query \T{X = [A, B, C], q(X)}, however, partially
binds \T{X} before the call to \T{q}, \ie, to all possible lists of length
three.  The final set of values for \T{X}
(if any) will be a subset of its initial set of values.

As with preconditions in Hoare logic, we can restrict the set of input
values of the free variables for which a program is defined (see
assumptions below).
}

\begin{figure}
\begin{center}
\begin{tabular}{rcl}
    $\Spec{P}$ & - & specification\\
    \{A\} & - & assumption \\
    $(S \lor T)$ & - & disjunction \\
    $(S \land T)$ & - & parallel conjunction \\
    $(S,T)$ & - & sequential conjunction \\
    $(\exists V \dot S)$ & - &  existential quantification \\
    $(\forall V \dot S)$ & - &  universal quantification \\
    $pc(U)$ & - & \mbox{procedure call} \\
\end{tabular}
\caption{Summary of commands in the wide-spectrum language}
\label{wide-spec-lang}
\end{center}
\end{figure}

\paragraph{Specifications.}
A specification $\Spec{P}$
constrains the instantiations of its free variables so that they satisfy
predicate $P$; it is the basic building block of programs in the wide-spectrum
language.
For example, the specification $\Spec{X
= 5 \lor X = 6}$ represents the set of instantiations $\{ 5,6 \}$ for
$X$.  
We define two special cases of specification commands:
\[
    \Fail \sdef \Spec{false} \\
    \True \sdef \Spec{true} \\
\]
The specification $\Fail$ is not satisfied by any instantiation of free
variables; 
it is like
Prolog's \T{fail}.  The specification $\True$ does nothing, \ie, 
does not constrain the instantiations; it is like
Prolog's \T{true}. 
Specification commands operate on any input instantiations, 
that is, their assumption is always $true$.

\paragraph{Assumptions.}
An assumption $\Ass{A}$, where $A$ is a predicate, 
acts as a precondition, and thus
restricts the input instantiations.
As such, it provides a context for a
program fragment.
For example, some program $S$
may require that an integer parameter be non-zero, which can be
expressed as ``$\Ass{X \neq 0}, S$''.
If the assumption does not hold, the program may abort.
Aborting includes program behaviour such as non-termination and abnormal
termination due to exceptions like division by zero,
as well as termination
with arbitrary results.  
We define the (worst possible) program $\Abort$:
\[
    \Abort \sdef \Ass{false}
\]
The program $\Abort$ is thus undefined for any input instantiations.

\paragraph{Program Operators.} The disjunction of two programs $(S \lor T)$
behaves similarly to logical disjunction.
The output instantiations of a disjunction is the union of the instantiations
of the two programs.
There are
two forms of conjunction: a parallel version $(S \land T)$, where $S$ and
$T$ are executed independently and the intersection of their
instantiations is formed on completion; and a sequential form $(S,T)$,
where $S$ is executed before $T$, and hence $T$ can rely on the context
established by $S$.

\paragraph{Quantifiers.}
The existential quantifier $(\exists V@S)$ generalises disjunction,
computing the union of the results of
$S$ for all possible values of $V$.
Similarly, the universal quantifier $(\forall V @ S)$ generalises
conjunction,
computing the intersection of the results of $S$ for all
possible values of $V$.

\paragraph{Procedure call.}
A procedure call is of the form $pc(U)$, where $pc$ is a procedure
and $U$ is a list of terms.

\subsection{Procedure definitions}

A summary of the syntax associated with procedures is given in
\reffig{wspl-procs}.

\begin{figure}
\begin{center}
\begin{tabular}{rcl}
    $V \prm S$ & - & \mbox{procedure} \\
    $\re p @ V \prm \C(p) \er$ & - & \mbox{recursive procedure} \\
    $id \sdef proc$ & - & \mbox{procedure definition}
\end{tabular}
\caption{Summary of procedure definitions}
\label{wspl-procs}
\end{center}
\end{figure}

\paragraph{Procedures.}
A (non-recursive) procedure is of the form $V \prm S$, where 
$V$ is a list of formal parameters and $S$ is the body of the
procedure (a command).  

\paragraph{Recursive procedure.}
A recursive procedure has the form $\re p @ V \prm \C(p) \er$.
Its body, $\C(p)$, encodes zero or more recursive
calls to $p$. 
To guarantee termination,
the actual parameters of the recursive calls must be less than 
the formal parameters ($V$)
according to some well-founded relation.

\paragraph{Procedure definition.}
A procedure definition is of the form $id \sdef proc$, where
$id$ is the name of
the procedure and $proc$ is a (recursive or non-recursive) procedure.

A distinguishing feature of the refinement calculus when compared to
most logic program synthesis schemes is the inclusion of assumptions.
This allows one to easily distinguish between what is
assumed by a program and what the program must establish.
This is useful when defining procedures;
often a procedure assumes the type of some of its parameters,
\eg, $\Ass{X \in list(\nat)}$.
This assumption may simplify the refinement --- without it some of the
desired properties of the parameter cannot be used.
Alternatively a procedure may be specified to \emph{establish} the type of
one of its parameters, by giving the type in a specification rather
than an assumption, \eg, $\Spec{X \in list(\nat)}$.
In logic programming terms,
in the case where a type is given in an assumption
the actual parameter to the procedure must be bound to a term of that
type.
The actual parameter must satisfy whatever
assumptions are made about it, or the procedure may abort.

\paragraph{Example.}

We may specify a procedure $reverse$ that relates a list with its reverse.
We assume list indices start at 1.
\[
    reverse \sdef (L,R) \prm \\
		\t1 \Ass{list(L)}, \\
        \t1 \Spec{list(R) \land \#L = \#R} \land \\
        \t1 \Spec{(\all i : 1..\#L @ L(i) = R((\#L - i)+1))}
\]

We have defined $reverse$ to be a procedure with formal
parameters $L$ and $R$.  Within the body of the definition, we assume that $L$
is a list, giving the type of $L$ as well as ensuring that $L$ must be
bound before a call to $reverse$.
The procedure is then required to establish that $R$ is a list of the same
size as $L$, and that the elements of $R$ are the same as those of $L$,
but in reverse order.

A more concrete implementation of the $reverse$ specification 
is given by the following recursive program\footnotemark.
\footnotetext{A refinement of the abstract $reverse$ definition to the
recursive version may be found in \citeN{Thesis}.}
{\Defn[Reverse of a list]
\label{reverse:defn}
\[
    reverse \refsto \re rev @ (L,R) \prm \\
        \t1 \Spec{L= \el \land R = \el} \lor \\
        \t1 (\exists H, T @ \Spec{L = [H|T]}, \\
            \t2 (\exists RT @ append(RT, [H], R) \land
                rev(T, RT))) \er\\
\]
}
We have a recursive block that uses the name $rev$ for recursive calls.
The body is a disjunction; the first disjunct is the base case where $L$ is
empty, and therefore $R$ is also empty.
The second disjunct is the recursive case, where $L$ is nonempty.
We reverse the tail of $L$ with the recursive call $rev(T, RT)$, and append 
the head of $L$, $H$, onto the end of $RT$
($append$ defines the relationship between three lists where the third
is the concatenation of the first two).



\COMMENT{
We may specify a procedure $length$ that relates a list with its length.
{\Defn List length \label{list:length}}
\[
	length \sdef (L,N) \prm \Ass{list(L)}, \Spec{\#L = N}
\]

We have defined $length$ to be a procedure with formal
parameters $L$ and $N$.  Within the body of the definition, we assume that $L$
is a list, giving us the type of $L$ as well as ensuring that $L$ must be
bound before a call to $length$.
The procedure is then required to establish that $N$ is equal to the size
of $L$ (implying that $N$ is a natural number).

The procedure $length$ may be implemented using recursion.
\[
    \re lng @ (L,N) \prm \\
        \t1 (\Spec{L = \el} , \Spec{0 = N}) \lor \\
        \t1 (\exists H,T @ \Spec{L = \HT}  , lng(T,N-1)) \\
	\er
\]

This procedure is a recursive block using the name $lng$ for recursive calls.
The body is a disjunction; the first disjunct is the base case where $L$ is
empty, and therefore $N$ is 0.
The second disjunct is the recursive case, where $L$ is nonempty, and a
recursive call is made to $lng$.
The derivation of the recursive procedure from the specification
is presented in \refsect{gen:example}.
}

\subsection{Refinement}


Program $S$ is refined by program $T$,
written $S \refsto T$, if $T$ aborts less often than $S$, and when 
$S$ does not abort, $T$ produces the same answers as $S$.
Program equivalence ($\refeq$) is defined as refinement in both
directions.

This definition of refinement does not allow the reduction of
nondeter\-minism that imperative refinement allows; in logic
programming we are interested in
$all$ possible solutions, and hence any
refinement must also return all of those solutions.

\subsection{Refinement laws}
\label{pred:reftlaws}

In this section we present some basic refinement laws\footnote{%
All refinement laws used in this paper
have been proved correct with respect to the semantics of the language
\cite{Thesis}.}%
.
Each law represents a refinement (synthesis/transformation) that may be made.
Where a law is divided into two parts by a horizontal line,
the part above the line is the proof
obligation that must be satisfied for the refinement below the line
to be applied.
For example, \reflaw{weaken:ass} allows an assumption $\Ass{A}$
to be refined to $\Ass{B}$, if $A$ entails $B$.
This corresponds to reducing the
conditions under which the program can abort.
\reflaw{equiv:specs:rc} allows
the characteristic predicate of a specification to be replaced with an
equivalent predicate.  This corresponds to maintaining the set of answers for
free variables.
These two laws embody the definition of refinement;
they are the main laws we use for manipulating predicates.
\begin{center}
\parbox{60mm}{
\ReftLaw[Weaken assumption]{
\label{weaken:ass}
$\t1
    \Rule{A \entails B}
    {\Ass{A} \refsto \Ass{B}}
$
}
} 
\parbox{60mm}{
\ReftLaw[Equivalent specifications]{
\label{equiv:specs:rc}
$\t1
    \Rule{P \equiv Q}
        {\Spec{P} \refeq \Spec{Q}}
$
}
}
\end{center}
An entailment $A \entails B$ holds if and only if 
$A \imp B$ holds for all possible values of the free variables in the
predicates $A$ and $B$.
The equivalence operator $\equiv$ is defined as
entailment in both directions.

\COMMENT{
\reflaw{ctx:ass:spec:refsto:rc}
demonstrates the effect of contextual refinement.  It is similar to
\reflaw{equiv:specs:rc}, however we may use the assumption $A$ in discharging
the proof obligation $P \iff Q$.  We discuss contextual refinement in more
detail in \refsect{context}. 
{\ReftLaw[Assumption in context]
\label{ctx:ass:spec:refsto:rc}}
\[  
    \Rule{A \entails (P \iff Q)}
        {\Ass{A}, \Spec{P} \refsto \Ass{A}, \Spec{Q}}
\]  
}

\COMMENT{
\reflaw{lift:disj} allows predicate disjunction to be lifted to program
disjunction when the disjunction appears inside a specification.  
Note that the operator `$\lor$' on the left is predicate disjunction while
on the right `$\lor$' is program disjunction.  
The context always determines which operator is meant, and the two
operators have similar properties.
Predicate
conjunction and quantification may also be lifted to their program counterparts
when they appear inside specifications.
{\ReftLaw[Lift disjunction]
\label{lift:disj}}
\[  
    \Spec{P \lor Q} \refeq \Spec{P} \lor \Spec{Q}
\]
}

\reflaw{mono:para}
is an example of a monotonicity law.  In general, a monotonicity law
states that the refinement of a component of a program refines the entire
program.
In this case,
if $S$ refines to $S'$ and $T$ refines to $T'$ then the
parallel conjunction $S \land T$ refines to $S' \land T'$.
Monotonicity holds for all the operators and both quantifiers in the
wide-spectrum language.
\begin{center}
\parbox{80mm}{
\ReftLaw[Monotonicity  of parallel conjunction]{
\label{mono:para}
$\t1
    \Rule{ S \refsto S';\ T \refsto T'}
         { S \land T \refsto S' \land T' }
$
}
}
\end{center}

\COMMENT{
\reflaw{dist:seq:disj:rc} may be used to distribute a sequential
conjunction through a disjunction.
\[
    S, (T \lor U) \refeq (S, T) \lor (S, U)
\]

\reflaw{parameterise:rc} may be used to
unfold a non-recursive parameterised command.
The expression $\replace{S}{V}{U}$ represents the command $S$ with all
occurrences of free variable $V$ replaced with term $U$.
\[
    \replace{S}{V}{U} \refeq (V \prm S)(U)
\]
}


\COMMENT{
A more complex refinement law is that for introducing recursion.

{\ReftLaw Recursion introduction. \label{rec:intro} 
Let $pc:PCmd$ be a parameterised command; $(\_ \prec \_): Term \rel
Term$ a well-founded relation; and
$id$ a fresh name; then }
\[
    \Rule{(\all Y:Term @ \Ass{Y \wflt X}, pc(Y) \refsto id(Y)) \imp
        pc(X) \refsto \C(id)}
    {pc \refsto (\re id @ X \prm \C(id) \er)}
\]
For notational convenience
we have made some technical simplifications, including removing reference to an
environment (a mapping from procedure names to procedures).
The full law is derived from the principle of well-founded induction
in \cite{reflp:01}.
In \reflawN{rec:intro} the parameterised command $pc$ is refined to a 
recursion block.
The body of the recursion block is a command
$\C(id)$ with free variable $X$, and which may contain
calls on $id$.
To apply the law we must discharge the proof obligation.
This is done by showing that $pc(X) \refsto \C(id)$, assuming that
\[
    (\all Y: Term @ \Ass{Y \wflt X},pc(Y) \refsto id(Y))
\]
We call this the inductive hypothesis.
It allows recursive calls as long as the parameter $Y$ is less than $X$
according to some well-founded relation $\wflt$,
ensuring that the
recursion will terminate.
$pc(X)$ is just the body of $pc$, which we call $S$
($S$ is a command with free variable $X$, \ie, $pc \sdef X \prm S$).

\renewcommand{\theenumi}{R-\roman{enumi}}

The law is used in the following steps:
\begin{enumerate}
\item
\label{ri-step1}
Focus on the body of $pc$, $S$.
\item
\label{ri-step2}
Refine $S$, possibly using the inductive hypothesis to introduce calls to
$id$.
\item
\label{ri-step3}
Call this refined program $\C(id)$.
\item
\label{ri-step4}
Then, the proof obligation for the law has been proved (by instantiating
$\C$),
and the original non-recursive procedure $pc$ has been refined to the
recursive procedure
$(\re id @ X \prm @ C(id))$.
\end{enumerate}
The second step will in general be complex,
involving user direction as with most
program derivations.
The other steps in the process are trivial, including the construction of the
inductive hypothesis, which is determined by the syntactic form of the
specification.
An example that uses \reflaw{rec:intro} 
is presented in \refsect{gen:example}.

\renewcommand{\theenumi}{\arabic{enumi}}
}

\COMMENT{
We may also unfold a call on a recursive block.  Inside the block is
a program $\C(p)$, which is a command with free variable $V$ and recursive
calls on
procedure $p$.  If the block is applied to actual
parameter $T$, it is equivalent to $\replace{\C(p')}{V}{T}$,
where $p'$ is the original
recursion block.
{\ReftLaw Unfold recursion \label{rec:unfold:rc}}
\[
    (\re p @ V \prm \C(p) \er)(T) \refeq 
        \replace{\C(\re p @ V \prm \C(p) \er)}{V}{T}
\]
\reflaw{rec:unfold:rc} may be reformulated as follows.
\[
    pc(T) \refeq \replace{\C(pc)}{V}{T} \\
    \mbox{where}\ pc \sdef \re p @ V \prm \C(p) \er
\]
For example, given the following definition,
\[
    nats \sdef \rec(nat, {\pcmd(V, {\Spec{V = 0} \lor 
        (\exists Y @ \Spec{V = s(Y)}, nat(Y))})})
\]
a procedure call $nats(T)$ refines to
\[
    \Spec{T = 0} \lor (\exists Y @ \Spec{T = s(Y)}, nats(Y))
\]
}

\COMMENT{
\subsection{Example refinement}

We present an example refinement using some of the laws presented above.  We
implicitly use monotonicity so that the refinement of a subprogram refines the
program as a whole.
\begin{derivation}
    \step{\Ass{X = 5}, \Spec{Y = X+10 \lor Y = X + 20}}
    \trans{\refeq}{\reflaw{lift:disj}}
    \step{\Ass{X = 5}, (\Spec{Y = X+10} \lor \Spec{Y = X + 20})}
    \trans{\refeq}{\reflaw{dist:seq:disj:rc}}
    \step{(\Ass{X = 5}, \Spec{Y = X+10}) \lor (\Ass{X = 5}, \Spec{Y = X + 20})}
    \trans{\refsto}{\reflaw{ctx:ass:spec:refsto:rc} \crosstwo}
    \step{(\Ass{X = 5}, \Spec{Y = 15}) \lor (\Ass{X = 5}, \Spec{Y = 25})}
    \trans{\refsto}{\reflaw{remove:ass} \crosstwo}
    \step{\Spec{Y = 15} \lor \Spec{Y = 25}}
\end{derivation}
        
We leave the refinement at this point, since it is a disjunction of primitive
specification commands.  The proof obligation for
the second last step, in the two applications of \reflaw{ctx:ass:spec:refsto:rc},       
was:    
\[  
    X = 5 \entails (Y = X+10 \iff Y = 15) \land \\
    X = 5 \entails (Y = X+20 \iff Y = 25) \\
\]

}

\COMMENT{
An example program written in the language is:
{\Defn Reverse of a list \label{reverse:defn}}
\[
    reverse \sdef (L,R) \prm \\
        \t1 \Spec{list(R) \land \#L = \#R} \land \\
        \t1 \Spec{list(L)} \land 
            \Spec{(\all i : 1..\#L @ L(i) = R((\#L - i)+1))}
\]

We may implement this using recursion:
\[
    reverse \refsto \re rev @ (L,R) \prm \\
        \t1 \Spec{L= \el \land R = \el} \lor \\
        \t1 (\exists H, T @ \Spec{L = [H|T]}, \\
            \t2 (\exists RT @ append(RT, [H], R) \land
                reverse(T, RT))) \er\\
\]
where $append$ defines the usual relationship between lists.
In a later section we show the refinement steps taken.
}

%% file: contexts.tex
\section{Contextual refinement}
\label{context}

During refinement we often focus on a component of a program
and refine it,
resulting in a refinement of the entire program, \ie, our
wide-spectrum language is monotonic with respect to refinement.  In
many situations the larger program can provide context that
assists in the
refinement of a component.
This context can be used, for instance, to
discharge proof obligations.
In this section we introduce a general notion of context to the calculus,
and demonstrate its use with the refinement of a list-reversal
procedure.
The approach taken is particularly useful when using a refinement tool, as
demonstrated in \citeN{Hemer:01}. 
The tool can manage the context, instead of the user
having to explicitly pass the context
around in the form of assumptions.

\COMMENT{
Consider the two programs below:
\begin{align}
    ~& \Ass{list(L)}, S \label{length:ass} \\
    ~& \Spec{list(L)}, S \label{length:seq} 
\end{align}
Both programs are fundamentally
different in the way in which they state $list(L)$, which in turn
impacts on the context provided for the refinement of $S$.
Program~\refeqn{length:ass} assumes $L$ is a list,
and will abort if this is not the case; the
sequential conjunction implies that, during the refinement of $S$,
one may assume $list(L)$ holds.
Similarly in program~\refeqn{length:seq},
sequential conjunction is used and provides a strong context for the
refinement, though the program will fail rather than abort
if $list(L)$ does not hold.
In this section we describe how sequential conjunction can be used to 
provide context.
}


\COMMENT{
Program~\refeqn{length:para} is different again, as $list(L)$ occurs in
parallel with the body of the program.
Initially we can use $list(L)$ in the refinement of
$S$.  Then, if the resulting
implementation guarantees $list(L)$, we
may eliminate $\Spec{list(L)}$
(saving redundant type-checks in the final code).
}

\subsection{Context in refinement laws}
\label{context:reft:laws}

Some laws, such as \reflaw{equiv:specs:rc} 
are ``stand-alone'' laws.
Its premise, $P \equiv Q$, requires that $P$ must be equivalent to $Q$,
regardless of the context in which it appears.
However, 
we may wish to reference the context in order to discharge this proof
obligation.
To do this, we introduce a generalised form of \reflawN{equiv:specs:rc}.
\[
    \Rule{A \entails (P \iff Q)}{\Ass{A}, \Spec{P} \refeq \Ass{A}, \Spec{Q}}
\]
This law allows
assumptions to be used in the proof that $P$ is equivalent to $Q$.
We say the specification $\Spec{P}$ has $A$ in context.
Since we often encounter laws where a refinement occurs with respect to some
context we introduce an abbreviation.
\[
	A \inctx S \refsto S' \ \sdef \ \Ass{A},S \refsto \Ass{A},S'
\]
This is similar to the notation used by \citeN{Nickson:97} for contextual
refinement of imperative programs.
Thus the generalised form of \reflawN{equiv:specs:rc} is written as 
\ReftLaw[Equivalent specifications w.r.t. context]{
\label{equiv:specs:wrt}
$\t1
    \Rule{A \entails (P \iff Q)}
        {A \inctx \Spec{P} \refeq \Spec{Q}} 
$
}


\COMMENT{
In the next section we introduce
a shorthand for doing this, to simplify laws and refinements.

\subsection{Predicate definition of refinement}
\label{ctx:entails}

A refinement of a program $S$ to $S'$ with a predicate $A$ in context
is given by
\[
    \Ass{A}, S \refsto \Ass{A}, S'
\]
We manipulate this expression.
\begin{derivation}
    \step{\Ass{A}, S \refsto \Ass{A}, S'}
    \trans{\iff}{definition}
    \step{(A \land ok.S) \entails
        ((A \land ok.S') \land (ef.S \iff ef.S'))}
    \trans{\iff}{rearrange}
    \step{A \entails (ok.S \imp (ok.S' \land (ef.S \iff ef.S')))}
\end{derivation}

The predicate on the right-hand side of the entailment corresponds to a
predicate version of refinement (\refdefn{pred:refsto}).
We define this using a new operator, $\refstar$.
\begin{equation}
    S \refstar T \hskip 2mm == \hskip 2mm 
        ok.S \imp (ok.T \land (ef.S \iff ef.T)) \label{pred:refstar}
\end{equation}

We may continue the above manipulation:
\begin{derivation}
    \step{A \entails (ok.S \imp (ok.S' \land (ef.S \iff ef.S')))}
    \trans{\iff}{from \refeqn{pred:refstar}}
    \step{A \entails (S \refstar S')}
\end{derivation}

We have reformulated the refinement of a program $S$ to $S'$
with assumption $A$ in
terms of predicate refinement between $S$ and $S'$, with $A$ acting as an
antecedent.  In practice, it can be used to discharge any proof obligations
arising during the refinement of $S$.

Since context will be represented using assumptions,
we introduce the following
shorthand for the expression $\Ass{A}, S \refsto \Ass{A}, S'$.
\[
    A \inctx S \refsto S'
\]
This is similar to the notation used in \cite{Nickson:97}.
The expressions $A \inctx S \refsto S'$ and
$A \entails S \refstar S'$ are equivalent, though we prefer to use
the former in presenting refinement laws since it uses the original version of
refinement.
}

The following law is similar to \reflaw{mono:para}, except that context for a
parallel conjunction is inherited by both conjuncts.
\ReftLaw[Monotonicity of parallel conjunction]{
\label{mono:para:ctx} 
$\t1
    \Rule{(A \inctx S \refsto S');\ (A \inctx T \refsto T')}
        {A \inctx S \land T \refsto S' \land T'} \hskip 5mm
$
}

To refine $S \land T$ in context $A$,
we may refine either of the conjuncts $S$ or $T$ 
using $A$ as the context.
There are similar contextual monotonicity laws for the other constructs in our
language.
Such laws allow the context to be passed around in a straightforward manner, 
and for this reason we do not explicitly mention the application of
such laws in refinements.



For a sequential conjunction $(S,T)$, command 
$S$ is executed before $T$, and hence
$S$ establishes a context for $T$.
For example, in the program $\Spec{X = 1},\Spec{Y = X + 1}$, the first
component establishes $X=1$, and this may be assumed when refining the
second component, \eg, the second component may be refined to $\Spec{Y=2}$.
\reflawN{context:seq:ass} gives the general rule when the first component is
an assumption, and \reflawN{context:seq:spec} when it is a specification.

\begin{center}
\parbox{60mm}{
\ReftLaw[Assumption in context]{
\label{context:seq:ass}
$\t1
    \Rule{A \land B \inctx (T \refsto T')}
        {A \inctx \Ass{B},T \refsto \Ass{B},T'}
$
}
}
\parbox{60mm}{
\ReftLaw[Specification in context]{
\label{context:seq:spec}
$\t1
    \Rule{A \land P \inctx (T \refsto T')}
        {A \inctx \Spec{P},T \refsto \Spec{P},T'}
$
}
}
\end{center}

Using \reflawN{context:seq:ass}
we may refine $T$ with $B$
in context in addition to $A$, and similarly 
with $P$ in \reflawN{context:seq:spec}.  
This information may be used to discharge proof
obligations in the refinement of $T$ to $T'$.

\COMMENT{
\subsection{Contextual refinement example}
\label{gen:example}

\setcounter{ctxCnt}{0}
\begin{figure}[p]
\begin{center}
\begin{derivation}
	\ContextBox{Actx}{(\all T,N' @
		\Ass{\#T<\#L}, \Ass{list(T)}, \Spec{\#T = N'} \refsto lng(T,N'))}
	\step{\Ass{list(L)}, \opening{\Spec{\#L = N}}}
    \trans{\refsto}{Window (\reflaw{context:seq:ass})}
    \begin{derivation}
        \Context{Bctx}{list(L)}
        \step{\Spec{\#L = N}}
        \trans{\refsto}{\reflaw{case:anal:seq} from \refcontext{Bctx}}
        \step{\Spec{L = \el}, \opening{\Spec{\#L = N}} \lor
            \Spec{(\exists H,T @ L = \HT)} , \Spec{\#L = N}}
        \trans{\refsto}{Window (\reflaw{context:seq:spec})}
            \begin{derivation}
                \Context{Cctx}{L = \el}
                \step{\Spec{\#L = N}}
                \parbox{20mm}{\newstep{Xstep}}\hskip-20mm%
                \trans{\refeq}{\reflaw{equiv:specs:wrt}
                        using \refcontext{Cctx}}
                \step{\Spec{0 = N}}
            \end{derivation}
        \step{(\Spec{L = \el}, \closing{\Spec{0 = N}}) \lor
            (\Spec{(\exists H,T @ L = \HT)}  , \Spec{\#L = N})}
        \trans{\refeq}{\reflaw{lift:exists},
            \reflaw{exists:extend:scope:seqL}}
        \step{(\Spec{L = \el} , \Spec{0 = N}) \lor
            (\exists H,T @ \Spec{L = \HT}  , \opening{\Spec{\#L = N}})}
        \trans{\refsto}{Window (\reflaw{context:seq:spec})}
            \begin{derivation}
                \Context{Dctx}{L = \HT}
                \step{\Spec{\#L = N}}
                \trans{\refeq}{\reflaw{equiv:specs:wrt} using \refcontext{Dctx}}
                \step{\Spec{\#\HT = N}}
                \trans{\refeq}{\reflaw{equiv:specs:wrt}}
                \step{\Spec{1 + \#T = N}}
                \parbox{20mm}{\newstep{Ystep}}\hskip -20mm
                \transbox{\refeq}{\reflaw{intro:ass}
                    using \refcontext{Bctx} and \refcontext{Dctx}, rearrange}
                \step{\Ass{list(T)}, \Spec{\#T = N-1}}
                \trans{\refeq}{\reflaw{intro:ass}
                    since $\#T < \#L$}
                \step{\Ass{\#T < \#L}, \Ass{list(T)}, \Spec{\#T = N-1}}
                \parbox{20mm}{\newstep{Zstep}}\hskip -20mm
                \transbox{\refsto}{inductive hypothesis (\refcontext{Actx})}
                \step{lng(T,N-1)}
            \end{derivation}
        \step{(\Spec{L = \el} , \Spec{0 = N}) \lor
            (\exists H,T @ \Spec{L = \HT}  , \closing{lng(T,N-1)})}
    \end{derivation}
    \step{\re lng @ (L,N) \prm}
        \step{\t1 \Ass{list(L)}, }
        \step{\t1 \closing{(\Spec{L = \el} , \Spec{0 = N}) \lor
            (\exists H,T @ \Spec{L = \HT}  , lng(T,N-1))} \er}
\end{derivation}
\end{center}
\caption{Refinement of $length$}
\label{length:seq:reft}
\end{figure}

In this section we show an example in which the context provided by sequential
conjunction is necessary to establish an assumption
to introduce a recursive call.
}

\COMMENT{
We first define the predicate $list$:
{\Defn ~\label{defn:list}}
\[
    list(L) == L = \el \lor (\exists H,T @ L = \HT \land list(T))
\]
We assume that the constant $\el$ and the
functor $[.|.]$ are contained in $Fun$ (see \ref{def-fun}).
We also assume the standard induction property on lists,
and therefore that our lists are finite.
}

\COMMENT{
Recall \refdefn{list:length} from \refpage{list:length} \ :
$length \sdef (L,N) \prm \Ass{list(L)}, \Spec{\#L = N}$.
We refine the definition into a recursive
procedure, and hence we use \reflaw{rec:intro}.
In this example we choose our well-founded relation to be the length of
the list parameter, which will be strictly decreasing.
The inductive hypothesis in the proof obligation for \reflaw{rec:intro} is
\begin{equation}
\label{length-ind-hyp}
	(\all T,N' @ \Ass{\#T < \#L},
		\Ass{list(T)}, \Spec{\#T = N'} \refsto lng(T,N'))
\end{equation}
where we have chosen $lng$ as the name for the recursive calls ($id$ from
\reflaw{rec:intro}).

\COMMENT{
\[
    length \refsto \re lng @ (L,N) \prm \\
        \t1 \Ass{(\all T,N' @ \Ass{\#T < \#L},
            \Ass{list(T)}, \Spec{\#T = N'} \refstar lng(T,N'))}, \\
        \t1 \Ass{list(L)}, \Spec{\#L = N} \er
\]
}

The refinement is shown in \reffig{length:seq:reft}.
We being the refinement on the body of $length$ (\refristep{1}).
Initially, the context, $A$, only contains the inductive hypothesis
\refeqn{length-ind-hyp}.
Each time we `window' on the right-hand side of a sequential
conjunction,
we add to the context (and subsequently remove from the context at the
corresponding `close window').  This is
justified by Laws~\ref{context:seq:ass} and~\ref{context:seq:spec}.  
The context can then be used to discharge
proof obligations.

For instance, in \refstep{Xstep} the proof obligation for using
\reflaw{equiv:specs:wrt} is $\#L = N \iff 0 = N$.  From \refcontext{Cctx} we
have $L = \el$, which entails $\#L = N \iff 0 = N$.
The proof obligation for \refstep{Ystep} is $list(T)$.
We use \refcontext{Dctx} as well as the `inherited' \refcontext{Bctx}, since
\[
    list(L) \land L = \HT \entails list(T)
\]
At \refstep{Zstep} we use the inductive hypothesis, \refcontext{Actx},
to introduce a recursive call (\refristep{2}).

After \refstep{Zstep} we have completed the refinement of the body of length
(the resulting command is $\C(id)$ of \refristep{3}).  
This completes the proof obligation for \reflaw{rec:intro}, and thus we have
refined the original parameterised command $length$ to
\[
    \re lng @ (L,N) \prm \\
        \t1 \Ass{list(L)},  \\
        \t1 (\Spec{L = \el} , \Spec{0 = N}) \lor \\
        \t1 (\exists H,T @ \Spec{L = \HT}  , lng(T,N-1)) \er
\]

}

%% file: datareft.tex

\COMMENT{
This section discusses data refinement within the logic programming
refinement calculus.  
Data refinement is a method for
replacing a data type within a program with another data type,
maintaining the original meaning of the program.  
Data refinement is simplified by contextual refinement.
In this section we focus on data refinement on a procedure level; this is
extended to groups of procedures in the next section.

Data refinement may be used in the following situations:
to replace an abstract specification type with an implementation type
(\eg, to represent a set using a list);
or to replace an implementation type with a more efficient type.
As an example of the latter, 
some list operations may be performed more efficiently by using a
difference list rather than a list (a difference list is
a pair of lists such that the second is
a suffix of the first).

Data refinement
provides a formal method for making these changes to a program.
Some of the important components of data refinement are discussed
below.  

{\bf Abstract and Concrete}\ \ A data type that is to be data-refined 
is called
$abstract$, while the target data type is called $concrete$.
Similarly, a variable that is an instance of either of these types is
referred to as an abstract or concrete variable.  Within general
program fragments abstract and concrete variables
are named $A$ and $C$, respectively.

{\bf Coupling invariant}\ \ A $coupling~invariant$ ($CI$) is a
predicate that relates the abstract and concrete variables.
For instance, a coupling invariant that relates sets to lists may
state that the elements in an (abstract) set are exactly the elements
contained in the range of
a (concrete) list.  Notationally, if the operator `ran'
returns the set of elements in the range of a list, this coupling
invariant may be expressed as $A = \ran C$.  

We distinguish three cases for logic data refinement:
\begin{enumerate}
\item {\em Parameter and interface refinement\/}: 
we data refine a parameter to a procedure, forcing a change in the
interface.
This case of data refinement
may be used when the coupling invariant
is established in the context of a call to the procedure.
\item {\em Parameter refinement\/}: the variable we are refining is a
parameter to the procedure we are interested in, but we want to keep
the old interface to the procedure.
\item {\em Local refinement\/}: the variable we are refining is internal
to the procedure we are interested in.
\end{enumerate}
In Sections~\ref{case1}-~\ref{case3}
we examine each of the cases in detail, 
and illustrate them with a running example.

\subsection{Parameter and interface refinement}
\label{case1}

The first case is data refinement of parameters.  This case is used
when at least one of the parameters to the procedure is to be
data-refined, and we also change the interface of the procedure.  
Unfortunately the change in interface will mean that
the new procedure cannot be used directly in place of the old, since
the types of the parameters will be inconsistent.  Thus this case of
data refinement will only be useful in contexts in which the
calling program establishes a coupling invariant.
In other words, the goal of this refinement is to derive a law of the
form:
\begin{equation}
\label{case1:reftrule}
		CI(A,C) \inctx {abs(A) \refsto conc(C)}
\end{equation}
where $abs(A)$ is the abstract program and $conc(C)$ is the concrete
implementation of it.
The easiest way to achieve this is to simply refine $abs(A)$ with
the coupling invariant in context.  When all references to $A$ have been
removed, the resultant program is called $conc(C)$; the data refinement is
complete, and the above law is a direct consequence.

In order for a program to use the concrete implementation rather than the
abstract, the coupling invariant must be 
established in the context of the call.  
We may then apply the instantiation of \refeqn{case1:reftrule} for the
particular program in question.

}

\subsection{Contextual data refinement}
\label{reverse:dr}

In this section we use contextual refinement to demonstrate
\emph{data refinement}, where a variable of
an \emph{abstract} type 
is replaced with one or more variables of a
\emph{concrete} type.
Data refinement may be used to replace a specification type with an
implementation type, or to improve the efficiency of a program.
The abstract and concrete types are related by a \emph{coupling invariant},
which is used to provide context for the data refinement.
As an example,
we show part of the refinement of 
the simple 
implementation of $reverse$ (\refdefn{reverse:defn})
on lists to a more efficient implementation using
difference lists (sometimes referred to as an accumulator implementation).
In \refsect{ctx-dr} we data refine $reverse$ assuming that the couping
invariant holds in context; in \refsect{int-dr} we complete the data
refinement by showing how the coupling invariant context can be established
efficiently and transparently.


\subsubsection{Coupling invariant in context}
\label{ctx-dr}
We refine a procedure call $reverse(L, R)$
in a context in which the list $R$ is represented by the difference list
$(DL1, DL2)$, \ie,
\begin{equation}
\label{rev-dr-ci}
	R \cat DL2 = DL1
\end{equation}
The operator `$\cat$' represents list concatenation, thus
$R$ is a prefix of $DL1$ and 
$DL2$ is a suffix of $DL1$.
When this relationship holds, $R = DL1 - DL2$ (interpreting `$-$' as list
difference).

\COMMENT{
This says that
the subtraction of $DL2$ from $DL1$
forms the list $R$.  We have the additional constraint $DL2 \suffix
DL1$,  which is true iff $DL1$ and $DL2$ are lists and $DL2$ is a 
a suffix of $DL1$.
}

We begin the refinement of $reverse(L,R)$, with the coupling invariant as an
assumption (the context for the refinement).
\[
	\Ass{R \cat DL2 = DL1}, reverse(L,R) 
\]
We expand the call $reverse(L,R)$ from \refdefn{reverse:defn}.
\[
	\Ass{R \cat DL2 = DL1}, \\
	\Spec{L= \el \land R = \el} \lor \\
	(\exists H, T @ \Spec{L = [H|T]}, \\
		\t1 (\exists RT @ append(RT, [H], R) \land rev(T, RT))) \\
\]

Because program disjunction is monotonic with respect to refinement
and the context of the disjunction is
inherited by its disjuncts, we may refine the first disjunct, $\Spec{L= \el
\land R = \el}$, with the coupling invariant in context.
Using \reflaw{equiv:specs:wrt} we rewrite $R = \el$ to $DL1 = DL2$, since
the context \refeqn{rev-dr-ci} implies they are equivalent expressions.
\[
	\Spec{L= \el \land DL1 = DL2 } 
\]
The details of the refinement of the second disjunct are
more complex, requiring the introduction of a recursive call.
We omit the details for brevity,
though the full refinement can be found in \citeN{Thesis}. 
The resulting recursive program is the usual difference list implementation
of $reverse$.
\[
    reversedl \sdef \re revdl @ (L,DL1,DL2) \prm \\
		\t1 \Spec{L= \el \land DL1 = DL2} \lor \\
		\t1 (\exists H, T @ \Spec{L = [H|T]} ,
				revdl(T, DL1, [H|DL2])) \er
\]
The refinement can be summarised by the following relation:
\begin{equation}
\begin{split}
\label{dr:rev:case1}
	~& {R \cat DL2 = DL1} \inctx 
		{reverse(L,R) \refsto reversedl(L,DL1,DL2)}
\end{split}
\end{equation}

\COMMENT{
\[
	reverse \sdef \re rev @ (L, R) \prm  \\
			\t3 \Spec{L= \el \land R = \el} \lor \\
			\t3 (\exists H, T @ \Spec{L = [H|T]}, \\ 
				\t4 (\exists RT @ rev(T, RT) \land append(RT, [H], R))) \er
\]
}

\COMMENT{
$reversedl$ is a more efficient list reversal procedure as it avoids the need
for list concatenation.
The refinement law \refeqn{dr:rev:case1} allows us to replace calls to
$reverse$ with calls to $reversedl$ when we are in a context
in which the second parameter is related to a difference list.
Such a context may occur when the parameter $R$ is a local variable.
}


\COMMENT{
\subsection{Parameter refinement}
\label{case2}

The second case of data refinement is similar to the first, in that a
parameter to the procedure we are interested in is data refined.  However,
in this case the abstract interface to the procedure is to be maintained.
For example, we may wish to replace the implementation of $reverse$ in
the previous section with one that uses $reversedl$, though we keep the
interface to $reverse$ the same.  The coupling invariant, instead of
being established by the calling program, is established internally to the
implementation of the call.  Note that this data refinement
will only be useful if the conversion from the abstract representation
to the concrete representation can be performed efficiently.

\COMMENT{
Consider the refinement of a call $abs(A)$, in a context in which there exists
a concrete value corresponding to the abstract variable $A$, \ie, $(\exists C @
CI(A,C))$.  We assume that case 1 data refinement has already been performed,
\ie, there exists a law similar in form to \refeqn{case1:reftrule}.
We take the following refinement steps:
\begin{derivation}
	\step{abs(A)}
	\trans{\refeq}{\reflaw{intro:spec2} from context}
	\step{\Spec{(\exists C @ CI(A,C))}, abs(A)}
	\trans{\refeq}{\reflaw{lift:exists}, \reflaw{exists:extend:scope:seqL}}
	\step{(\exists C @ \Spec{CI(A,C)}, abs(A))}
	\trans{\refeq}{law of the form \refeqn{case1:reftrule}}
	\step{(\exists C @ \Spec{CI(A,C)}, conc(C))}
\end{derivation}

In this context, the abstract call has been replaced by a program that calls
the concrete procedure.  To keep the distinction clear between the original,
abstract implementation $abs$, the concrete implementation $conc(C)$,
and the new implementation of $abs$ using the concrete procedure, we name the
latter $absnew$.  
}
The outcome of this case of data refinement is a law 
of the form:
\begin{equation}
\label{case2:reftrule}
	{(\exists C @ CI(A,C))} \inctx
		 {abs(A) \refsto absnew(A)}
\end{equation}
where
\[
	absnew \sdef A \prm (\exists C @ \Spec{CI(A,C)}, conc(C))
\]
We assume that case 1 data refinement has been performed, \ie, the following
has been proved.
\[
	{CI(A,C)} \inctx {abs(A) \refsto conc(C)}
\]
	
In practice, the programmer would simply replace the implementation of $abs$
with $absnew$, avoiding the need for both the definition of a new procedure 
and the refinement of any call to $abs$.  For clarity, we distinguish the two
implementations by name.

}

\subsubsection{Data refinement by establishing context}
\label{rev-internal-dr}
\label{int-dr}

In the previous section a context was given that allows
calls to $reverse$ to be replaced
with calls to the more efficient procedure $reversedl$.
However establishing this context in arbitrarily large and complex programs
may not be feasible.
In this section we show how the problem can be avoided by 
implementing $reverse$ in terms of $reversedl$. 


We start by choosing a stronger coupling invariant than 
\refeqn{rev-dr-ci}, in which $DL1$ is equal to $R$ and
$DL2$ is the empty list.
\begin{equation}
\label{ci:strong}
	R = DL1 \land DL2 = \el 
\end{equation}
Hence we may deduce
$reverse(L,R) \refsto reversedl(L, R, \el)$
because \refeqn{ci:strong} implies the premise 
of \refeqn{dr:rev:case1}.
This is a valid refinement in any context.
Of course, in a program that makes many calls to
$reverse$, we may hide this change by implementing
the body of $reverse$ as just a call to $reversedl(L, R, \el)$.
The (new) body of $reverse$ provides the context of \refeqn{ci:strong}
locally, avoiding the need for the calling program to establish the context.

The above refinements are examples of \emph{data refinement} on procedures.
In the next section we consider data refinement on groups of 
procedures that operate on a common data type.

\COMMENT{
In this case $(\exists C @ CI(A,C))$ is equivalent to $list(C)$.
Hence we deduce that, given \refeqn{dr:rev:case1},
\begin{equation}
\label{dr:rev:case2}
	{list(R)} \inctx
	{reverse(L,R) \refsto reversenew(L, R)}
\end{equation}
where
\[
	reversenew \sdef (L,R) \prm \\
		\t1 (\exists DL1, DL2 @ \Spec{R = DL1 \land DL2 = \el \land list(R)}, \\
		\t2 reversedl(L,DL1, DL2))
\]

We may simplify $reversenew$ to just $reversedl(L,R, \el)$.

}

\COMMENT{
\subsection{Local refinement}
\label{case3}

The final case of data refinement is performed on a local variable rather
than a parameter.  This case is most like imperative data refinement, since
there is no need to alter the procedure's interface.

For this type of data refinement, we simply refine the body of the
procedure we are interested in without reference to any context.
We assume that we have a program (fragment) scoped by an existentially
quantified abstract variable that has its type specified by a parallel
specification, i.e.,
\[
	(\exists A @ \Spec{type(A)} \land S)
\]
We introduce the concrete variable and coupling invariant
(\emph{augmentation}),
then manipulate the program so that the abstract variable is redundant.
We remove the redundant variable (\emph{diminution}).

Consider a procedure $palindrome(L, P)$ that holds if the (even length)
list $P$ is the
palindrome formed by appending the reverse of list $L$ to itself.
\[
	palindrome \sdef (L, P) \prm \\
		\t1 \Ass{list(L)} ,  (\exists R @ \Spec{list(R)} \land 
				reverse(L, R) \land append(L, R, P))
\]

We have an internal variable $R$ of type list, which is used in a call to 
$reverse$.
We replace $R$ with a difference list, we can make
a more efficient call to $reversedl$.  


\COMMENT{
Before progressing, we look at our proof obligation for 
\reflaw{dr:augment} using the above coupling invariant.
\begin{derivation}
	\step{(\exists DL1,DL2 @ DL1=R \land DL2=\el \land list(R))}
	\trans{\iff}{one point on $DL1$ and $DL2$}
	\step{list(R)}
\end{derivation}

The proof obligations for
$diminish$ \reflaw{dr:diminish} are:
\begin{derivation}
	\step{(\exists R @ DL1=R \land DL2=\el \land list(R))}
	\trans{\iff}{one-point rule on $R$}
	\step{DL2=\el \land list(DL1)}
\end{derivation}
and
\begin{derivation}
	\step{(\exists DL1, DL2 @ DL2=\el \land list(DL1))}
	\trans{\iff}{true}
\end{derivation}
}

The resulting implementation of $palindrome$, which uses the more efficient
procedure $reversedl$, is:
\[
	(\exists DL1 @ \Spec{list(DL1)} \land \\
		\t1 (reversedl(L,DL1,\el) \land append(L,DL1,P)))
\]

\COMMENT{
The resultant program is a more efficient implementation than the
original, since we use difference lists to calculate the reverse of
$L$.  
}

Any procedure that calls $palindrome$ does not need to change,
since although the body of palindrome has been refined, the interface
has remained untouched.

\COMMENT{
Note that in this instance there is the option to instead use
\refreft{dr:rev:case2} directly, \ie,
\begin{derivation}
	\step{(\exists R @ \Spec{list(R)} \land reverse(L,R) \land 
			append(L,R,P))}
	\trans{\refsto}{\reflaw{para:to:seq}}
	\step{(\exists R @ \Spec{list(R)} , (reverse(L,R) \land 
			append(L,R,P)))}
	\trans{\refsto}{\refreft{dr:rev:case2}}
	\step{(\exists R @ \Spec{list(R)} , (reversedl(L,R,\el) \land 
			append(L,R,P)))}
\end{derivation}
Subject to renaming the two programs are equivalent.  
}


\COMMENT{
The laws that form the basis of cases 1 and 2 of data refinement express 
the relationship between the abstract and concrete procedures.  However, the
proof obligations for these rules imply that a strong context exists at the
point of the call.  For instance, in \refreft{dr:rev:case2}, there is the
hypothesis $list(R)$.  However,
in logic programming terms, the parameter $R$ to reverse
is used as an output, \ie, $L$ is ground before the call and the procedure
produces a non-variable binding for $R$.  In the formulation of
\refreft{dr:rev:case2}, $R$
must be at least partially ground before the data refinement can take place.
This resulted in added complication when refining from sequential conjunction
to parallel conjunction in \reflaw{dr:diminish}.

In the next section we extend data refinement to consider groups of related
procedures, rather than individual procedures, and also the structure of the
programs in which they appear.  By tailoring laws to certain program
structures, we can eliminate the need for explicit assumptions about some
variables; for instance, using techniques in the next section, it is not
necessary to have make the specification $\Spec{list(R)}$ explicit
in the definition of
$palindrome$ in order to perform the refinement.
}

\COMMENT{
An important difference between imperative data refinement and logic
programming data refinement is that in the latter
the variables to be data refined cannot
be `hidden'.  We also find that the data refinements
presented here are 
defined with respect to standard refinement.
In imperative data refinement the data
refinement process is transformational \cite{Morgan:94},
and does not directly involve the definition of standard refinement.  
If the transformation follows
certain guidelines, then a program that uses the
abstract module is refined by a similar program that uses the concrete
module.  In the next section we take a similar approach, by introducing
the notion of modules in the wide-spectrum language, and extending the
data refinement process in this section from individual procedures to
groups of procedures.
}

}

%% file: modreft.tex
\section{Modular logic program refinement}
\label{modreft}

In this section we introduce the notion of a $module$,
which is a group of procedures that operate on a
common data type.
By making some assumptions about the context in which an abstract
module may be used,
we may allow a more efficient module to be used
in its place.



\COMMENT{
We distinguish abstract and concrete
module specifications.  For instance, an abstract module with 
a set data type
might include procedures to add an element to a set and to check
set membership.  We would like to be able to replace the calls to the
set module with calls to a module that implements the same operations,
except on an implementation type, \eg, Prolog lists.  
By considering programs that respect the intended modes of the module
procedures, 
we can derive
efficient concrete representations of abstract modules.
}

\COMMENT{
\refsect{module:specs} introduces modules into the wide-spectrum language.
In \refsect{sect:using:modules} we describe a structural form that
procedures that use a module must obey.
In \refsect{module:reft:sect} we prove the main module refinement theorem,
which describes a set of conditions that allows one module to be used in
place of another, resulting in a refinement of the calling program.
In \refsect{mr-example} we show as an example an array-based implementation of
a partial function.
}

\subsection{Module specifications}
\label{module:specs}

As with modules in logic programming languages such as Mercury
\cite{Mercury:95} and \Goedel\ \cite{Godel:94}, 
modules in the wide-spectrum language are collections of procedures that
operate on a common data type.  The data type is intended to be
$opaque$, that is, the
implementation of the type is hidden, and variables of that type may only be
manipulated via the procedures of the module.


We split the opaque parameters of a module procedure
into two categories, \emph{input} and \emph{output},
which correspond with the logic programming \emph{modes} ``ground'' and
``var'' (unbound), respectively.
Upon a procedure call,
opaque inputs must already have been instantiated to the module type and
opaque outputs must be uninstantiated.
In addition, procedures may have a
set of \emph{regular}, \ie, non-opaque, parameters.

\reffig{module:pfun} defines a module $\PFunName$
that declares operations on a type $\PFun$. 
The type $\PFun$ is a partial function
from elements of its domain
type $\sigma$ to elements of its range type $\tau$, written
$\sigma \pfun \tau$.
A function may be modeled as a set of pairs.
A partial function is a function that may be undefined for some elements of
its domain, as distinct from a total function which maps every element of
its domain to some value.
We have left the actual types for $\sigma$ and $\tau$
unspecified since none of the operations depend on these types
(though later we will assume that a hash function exists for $\sigma$) ---
we can therefore consider $\PFunName$ to be polymorphic.
Within the module
the type signature of each procedure is declared.
Opaque inputs have an assumption about their type
and the
specification of each procedure guarantees that the opaque outputs
are instantiated to be of the opaque type.
Opaque inputs and outputs are subscripted with $i$
and $o$, respectively.  
The parameters of type $\sigma$ and $\tau$ ($K$ and $V$) are
regular parameters.

In the definition of $update$, the symbol `$\oplus$' stands for function
override; the function $f \oplus g$ is the same as function $f$,
except with all elements in the domain of $g$ mapped according to $g$.
Therefore, $F \oplus \SKV$ is the same as $F$ but with $K$ mapped to
$V$ instead of $F(K)$.
In the definition of $remove$ we use domain subtraction `$\dsub$';
the function $\{K\} \dsub F$ is the same as $F$, except $K$ is no
longer in the domain.

\begin{figure}
\[
	\Module\ \PFunName \\
	\mtype\ \PFun \sdef \sigma \pfun \tau \\
	\also 
	\begin{array}{rl}
	init: & F': \PFun_o \\
	update: & K: \sigma, V: \tau, F: \PFun_i, F': \PFun_o \\
	access: & K: \sigma, F: \PFun_i, V: \tau \\
	remove: & K: \sigma, F: \PFun_i, F': \PFun_o \\
	\end{array}
	\also
	\begin{array}{rl}
	init \sdef & F' \prm \Spec{F' = \es} \\
	update \sdef & (K, V, F, F') \prm 
		\Ass{F \in \PFun \land K \in \sigma \land V \in \tau}, 
		\Spec{F' = F \oplus \SKV} \\
	access \sdef & (K, F, V) \prm 
		\Ass{F \in \PFun \land K \in \sigma},
		\Spec{K \in \dom(F) \land V = F(K)} \\
	remove \sdef & (K, F, F') \prm 
		\Ass{F \in \PFun \land K \in \sigma}, 
		\Spec{F' = \{K\} \dsub F}  \\
	\end{array} \\
	\also
	\End
\]
\caption{Abstract partial function module}
\label{module:pfun}
\end{figure}

Following the data type terminology of
\citeN{Liskov:86}, 
a procedure with no opaque inputs is referred to as an 
$initialisation$ procedure;
for example, $init$ is an initialisation procedure 
which instantiates the opaque output $F'$ to the empty function
(represented by the empty set of pairs).  
A procedure with no opaque outputs is referred to as an $observer$;
for example, $access$ is an observer that fails if the regular parameter $K$
is not in the domain of the opaque input function $F$, and instantiates
the regular parameter $V$ to $F(K)$ otherwise.
A procedure with both opaque inputs and outputs
is called a $constructor$; for example, the procedure $update$ has an opaque
output $F'$, which is the opaque input $F$ updated by the pair $(K,V)$.
A constructor can be likened to updating the state in an
imperative module.  Note that $init$, $update$, and $remove$ all
guarantee that their opaque
output is an element of $\PFun$.

\subsection{Using modules}
\label{sect:using:modules}

Our intuition is that a module is to be used opaquely
in the construction and maintenance of
some data structure throughout multiple procedure calls.  
We therefore
consider programs
whose procedure calls are ordered so that
the intended modes
of the opaque inputs and outputs are satisfied, and the variables used as
opaque inputs and outputs are local to the program.
For instance,
consider the following program that uses the module $\PFunName$.
It inserts the pairs $(a,2)$ and $(b,1)$ into a function and accesses the
value for $a$.
\begin{equation}
\begin{split}
\label{contains}
	~& (\exists F @ init(F), (\exists F' @ update(a, 2,F,F'), \\
		 ~& \t1(\exists F'' @ update(b, 1,F',F''), access(a,F'',X))))
\end{split}
\end{equation}
The use of sequential conjunction reflects the notion of the changing
state and also allows the assumptions of the later calls to be satisfied.
Initially, $F$ is instantiated to the empty function.  The two calls to
$update$ update $F$ to $F'$ and then to $F''$.  Overall,
the only variable we are interested in is $X$ --- the 
opaque parameters
are local because they are existentially quantified when they are
used as an output.
By only dealing with programs of this form,
we can use contextual information to derive more efficient
implementations of the module.  

To formalise this notion,
we say a program is in \emph{\opqf} with respect to a module $\cal{M}$ if, 
for all
procedure calls $p(V, I, O)$ where $p$ is in $\cal{M}$ and 
$V$ stands for the regular parameters, 
the opaque inputs $I$ are bound and the opaque outputs $O$
are not bound before the call.
Also, the opaque variables must not be used except by procedures in $\cal{M}$.
We first define \oopqf, which is a generalisation of \opqf.

\begin{Defn}[\Oopqf]  
\label{oopqf:form}
We say a program is in \emph{\oopqf} w.r.t. a module $\cal{M}$
and a set of free opaque variables $IV$
if it is in one of the following forms:
\begin{enumerate}
\item
\label{basecase}
a program fragment that does not rely on the opaque variables in $IV$ nor
make calls on any of the procedures in $\cal{M}$; 
\item
\label{disj}
a program of one of the following forms, 
\[
	\C_1 \lor \C_2 \\
	\C_1 \land \C_2 \\
	\C_1 , \C_2 \\
	(\exists V @ \C_1) \\
	(\all V @ \C_1) \\
\]
where $\C_1$
and $\C_2$ are subcomponents that are in \oopqf\ w.r.t. $\cal{M}$ and
$IV$, and $V$ is a regular (non-opaque) variable; 
or,
\item
\label{opq}
a program of the form 
\[
	(\exists O @ p(V,I,O), \C)
\]
where $p$ is a procedure in $\cal{M}$ and
$V$, $I$, and $O$ are the 
regular, opaque input, and opaque output parameters,
respectively, of $p$.
The opaque inputs $I$ must be a subset of $IV$.
The component $\C$ must be in \oopqf\ w.r.t. $\cal{M}$ and
the set $IV \union \{O\}$.
When $p$ has no outputs, \ie, it is an observer, there are no quantified
variables and the corresponding form is just $p(V, I), \C$.
\end{enumerate} 
\end{Defn}

\begin{Defn}[\Opqf] 
\label{pgm:form}
We say a program is in \emph{\opqf} w.r.t. a module $\cal{M}$
if it is in \oopqf\ w.r.t. $\cal{M}$
and contains no free opaque variables.  
\end{Defn}

Because logic programs do not typically have ``state'', we must pass 
the opaque parameters explicitly, and hence in some sense the
implementation details are exposed.
However, programs that are in \opqf\ are restricted to only using the
opaque type and variables via the procedures of the module.
This ensures the module is used as intended, \ie, with the type
being opaque.

Since the type is opaque,
a program in \opqf\ is amenable to syntactic
simplification that hides the opaque variables.  
The opaque variables are locally quantified, and
typically appear as input/output pairs, thus
we can adopt a shorthand similar 
to that of definite clause grammars (DCGs) in Prolog (and other logic
programming languages).
For instance, we could write program \refeqn{contains}
thus:
\begin{equation}
\begin{split}
\label{contains-simp}
	~& init, update(a, 2), update(b, 1), access(a,X)
\end{split}
\end{equation}
At each call to a procedure from the module, from form~\ref{opq} in
\refdefn{oopqf:form} 
we can immediately identify
that a new output opaque variable must be quantified (except in the case of
the observer, $access$), and fill in the
in/output parameters of the procedure call appropriately (resulting in
program \refeqn{contains}).
However, syntactic simplifications like this restrict expressiveness. 
For instance, by hiding the state we have no easy way of having two
instances of the state active at one time (imperative languages without
opaque types also have this problem). 
For instance, the shorthand notation cannot be used to simplify the
following program, which has two different partial functions $G$ and
$H$, containing $(b,1)$ and $(b,2)$, respectively.
\[
	(\exists F @ init(F),  
		 (\exists G @ update(b, 1,F,G), 
		 (\exists H @ update(b, 2,F,H), \hdots))
\]
For this reason we use the more general notation in which opaque
variables are explicit.

\COMMENT{
{\Defn \label{oopqf:form}
We say a program is in \emph{\oopqf} w.r.t. a module $\cal{M}$
if it is of the following structure, 
where $\C$, $\C_1$
and $\C_2$ are subcomponents that are also in \oopqf\ w.r.t. $\cal{M}$.
Programs in \oopqf\ may have a set of free opaque input variables, $IV$.
\begin{enumerate}
\item
\label{basecase}
a program fragment that does not manipulate the opaque variables in $IV$; 
\item
\label{disj}
a wide-spectrum program of one of the following forms, 
\[
	\C_1 \lor \C_2 \\
	\C_1 \land \C_2 \\
	\C_1 , \C_2 \\
	(\exists V @ \C) \\
	(\all V @ \C) \\
\]
where $V$ is a regular (non-opaque) variable, 
and;
\item
\label{opq}
a sequential conjunction of a call to a procedure, $p$, from
$\cal{M}$,
and a component, $\C$, 
with 
the opaque outputs, if any, existentially quantified,
\ie, $(\exists O @ p(V,I,O), \C)$.
$V$, $I$, and $O$ are the 
regular, opaque input, and opaque output parameters,
respectively, of $p$.  
The input parameters are subsets of $IV$.
\end{enumerate} }

In \refform{opq} we use the notation $p(V,I,O)$ as a short-hand for 
$p(V_1..V_i, I_1..I_j,$ $O_1..O_k)$ (where $i, j, k \geq 0$).  
The use of sequential conjunction gives an ordering to the
procedure calls that satisfies the intended modes, 
which helps with reasoning about the values of the
opaque parameters.

Each subcomponent may contain references to 
the set of opaque inputs $IV$.
These variables have been
quantified at an outer level (via \refform{opq}), 
but not in the component itself.  
In example \refeqn{contains}, the
call to $access$ has the variables $F$, $F'$ and $F''$ in scope,
although only $F''$ is used.
In \refform{opq}
the subcomponent, $\C$, includes the opaque outputs, $O$, of $p$ as 
variables that it may use as opaque inputs, along with the variables
inherited from $IV$.
For instance, in \refeqn{contains}, the program fragment 
$(\exists F'' @ update(b,1,F',F''), access(a,F'',X))$,
which is an instance of \refform{opq}, 
has $F$ and $F'$ as possible opaque inputs
(\ie, $IV = \{F,F'\}$), while the subprogram
$access(a,F'',X)$ has, in addition to these, $F''$.

We define a top-level program as a program that obeys the structure of
\refdefn{oopqf:form}, but contains no free opaque variables.
{\Defn \label{pgm:form}
We say a program is in \emph{\opqf} w.r.t. a module $\cal{M}$
if it is in \oopqf\ w.r.t. $\cal{M}$
and contains no free opaque variables.  }

An example of such a program is \refeqn{contains}.
We do not allow free opaque variable at the topmost level, \ie, the user level,
as the answers presented to the user would be different for different
implementation types used.
}

\COMMENT{
\subsection{Discussion}
\label{moduse:discussion}

In practice it would be useful to introduce some abbreviations to our
module template.
For instance, one could abbreviate the specifications in \reffig{module:pfun}
by omitting the explicit assumptions about the types of the opaque
inputs and using the convention that an
opaque input implicitly has its type assumption associated with it.
For example, $access$ would be abbreviated to
\[
	\Ass{K \in \sigma}, \Spec{K \in \dom F \land V = F(K)}
\]
Since we require the full definition of a module procedure for the purposes
of module refinement, we do not make such abbreviations in this
paper.

There are some differences between our approach
and the modules of, for example, Morgan \cite{Morgan:94},
stemming mainly from the differences between the imperative and logic
programming paradigms.  
Most noticeably,
Morgan's modules are used to encapsulate a state which is completely hidden
to an external user of the module.
Taking this approach, only one instance of a module's data type may be used by
a program.
Taking the opaque data type approach, however, allows more than
one instance of the module data type to appear in a program.
Our framework also allows
multiple initialisation procedures, and regular parameters to be included in
an initialisation (we could initialise a function to be a singleton set
containing the pair $(K,V)$);  the disadvantage is that the
initialisations must be called explicitly.  
Bancroft \cite{Bancroft:97} presents a framework for refinement of
imperative programming modules that define opaque types similar to ours.

We have stated that the
only procedures that may manipulate opaque-type variables are those
provided by the module.
However, in practice,
opaque parameters may be passed via intermediate procedure calls.
These intermediate procedure
calls must ensure the program as a whole remains
in \opqf,
given the application of \reflaw{parameterise} to the intermediate
procedure call.
}

\COMMENT{
\subsection{Equality}
\label{equality-discussion}

Our restrictions on programs using the module forbid the direct use of
equality on opaque-type variables.
This is because, for example, two abstract sets, $S$ and $S'$, may have the
same set value, but the concrete representations of the two variables
as lists, $L$ and $L'$, can be different lists
(containing the same elements in different orders).
If an equality operation is required on sets,
it should be provided explicitly via a module procedure,
which in the concrete list implementation must check whether the two
lists contain the same elements (or make use of a unique ordered
list to represent the set and use equality on lists).

Related to this,
each actual opaque output parameter must be a distinct
uninstantiated variable, that is, a variable may only be used as an opaque
output parameter once.
This is
because the use of an instantiated variable as an output encodes an
equality, \eg, if $S$ is instantiated then $p(S)$ is equivalent to
$(\exists S' @ p(S'), \Spec{S' = S})$.
For example, consider the following sequence of operations using the
set module.
\[
  \exists S @
    init(S) \sand
    \exists S' @ add(1,S,S') \sand
    \exists S'' @ add(2,S',S'') \sand \\
    \exists T @
    init(T) \sand
    \exists T' @ add(2,T,T') \sand
    add(1,T',S'')
\]
Note that the last parameter of the last call to $add$, $S''$,
is the same as the last parameter of the second call to $add$.
If we interpret this program as operations on sets we expect it to
succeed,
but an implementation using lists may fail because, although
the lists {\tt [1,2]} and {\tt [2,1]} have the same sets of elements,
they are different lists.
One could remove the restriction that outputs should be uninstantiated
by insisting on a unique concrete representation of each abstract
value,
however, the restriction on outputs allows a wider range of
implementations of modules.
The restriction may always be avoided by explicit use of
a set equality procedure provided by the module, \eg, the 
last call $add(1,T',S'')$
above could be replaced by
\[
    (\exists T'' @ add(1, T', T''), equals(S'',T''))
\]
assuming that the procedure $equals$ models set equality.
}


\subsection{Module refinement}
\label{module:reft:sect}

In general, we say a module $\cal{M}$ is refined by a module $\cal{M^+}$
if, for all possible programs $S$ using calls to $\cal{M}$,
$S$ is refined by the program $S^+$ obtained by
replacing all calls to the procedures of $\cal{M}$ by 
calls to the corresponding procedures of $\cal{M^+}$.
In this section we consider a law for module refinement 
(\refth{th:mod:reft})
that can be used only if
the programs using the module are in
\opqf\ (\refdefn{pgm:form}).

Consider the $\PFunName$ module defined in \reffig{module:pfun}.
A program that uses it,
\eg,
\refeqnContains, is not directly implementable, since the module
uses the abstract partial function type which is not part of the 
implementation language.  We would like to replace the
calls to $init$, $update$, $remove$, and $access$
from the $\PFunName$ module with corresponding
calls on a module that implements the operations on an implementation
data type.
Of course, replacing the references to the
$\PFunName$ module with references to the implementation module must 
result in a refinement of the program in question.
The following is our theorem for module refinement.
As with the data refinement example in \refsect{reverse:dr}, we require a
coupling invariant ($CI$) to relate the abstract and concrete types.

\COMMENT{
In general, this approach is not powerful enough to derive the efficient
implementations we want for module refinement.
The coupling invariant may introduce nondeterminism that is not
required in the implementation, by introducing multiple
concrete representations of an abstract value.
}

\begin{Theorem}[Module Refinement]  
\label{th:mod:reft}
Assume the following:
modules $\cal{M}$ and $\cal{M^+}$, with associated opaque types
$\Sigma$ and $\Sigma^+$, respectively; 
a coupling invariant $CI$, that relates the types $\Sigma$ and $\Sigma^+$;
and all corresponding pairs of procedures $p$ and $p^+$ 
from $\cal{M}$ and $\cal{M^+}$, respectively, 
satisfy \refcond{combined}, below, using $CI$.
Then
a program, $\C$, which is in \opqf\ w.r.t.
$\cal{M}$, is refined by the program $\C^+$, which is structurally the same
as $\C$ except with procedure calls to module $\cal{M}$
replaced by corresponding procedures calls to module $\cal{M^+}$.
\end{Theorem}
\Proof
The theorem is proved by structural induction over programs in
\oopqf.  A detailed proof can be found in \cite{Thesis}; it is a
generalised version of the proof in \cite{modreft:lopstr00}.\tqed

Consider the abstract and concrete procedures $p$ and $p^+$
which are defined as follows.
\[
	p \sdef (V, I, O) \prm \Ass{A}, \Spec{P} \\
	p^+ \sdef (V, I^+, O^+) \prm \Ass{A^+}, \Spec{P^+}
\]
The variables in $I$ and $O$ are of the abstract opaque type $\Sigma$, 
and similarly the variables in 
$I^+$ and $O^+$ are of the concrete opaque type $\Sigma^+$.
The regular variables, $V$, may be of any other type.
The free variables of the assumption $\Ass{A}$ are restricted to
$V$ and $I$, 
and the free variables of
the specification $\Spec{P}$ are restricted to $V$, $I$ and $O$.
Corresponding restrictions apply to $\Ass{A^+}$ and $\Spec{P^+}$.
The following predicate describes the conditions that must hold between 
procedures $p$ and $p^+$ 
with respect to the coupling invariant $CI$.

\begin{ReftCond} 
\label{combined}
\begin{align}
    CI(I,&I^+) \land A \entails \label{cond1:premise} \\
        & A^+ \land \label{proofob:noabort:p+} \\
        & (P \imp (\exists O^+ @ P^+ \land CI(O,O^+))) \land 
			\label{proofob:step1} \\
        & (P^+ \imp (\exists O @ P \land CI(O,O^+))) 
			\label{proofob:step5} 
\end{align}
\end{ReftCond} 

This condition states that, assuming the inputs are related by the 
coupling invariant and the abstract assumption $A$ holds
\refeqn{cond1:premise}: the concrete
assumption holds \refeqn{proofob:noabort:p+}; every abstract answer has a
corresponding answer in the concrete implementation \refeqn{proofob:step1};
and every concrete answer is related to an answer in the abstract procedure
\refeqn{proofob:step5}.

\subsection{Example}
\label{mr-example}

\paragraph{Concrete type.}
For our implementation of the $\PFunName$ module,
we assume
the existence of an injection\footnotemark, $hash$,
that uniquely maps elements of type $\sigma$ to a natural number
in the range $0..N-1$.
\footnotetext{
By requiring $hash$ to be an injection we are assuming that no two keys
will map to the same natural number and hence avoid the problem of
clashes.
A more general approach that handles clashes is possible, but would
complicate the presentation.
}
With this assumption,
we may implement a partial function as an array,
the indices of which are the hashed values of $\sigma$.
In other words, the array acts as a hash table.
We define the type $\Hash$ as an array of size $N$, the elements of which
are either the range type $\tau$ or the special element $\Null$
(not an element of $\tau$).
\[
	\Hash \sdef (0..N-1) \fun (\tau \union \{\Null\}) 
\]
The symbol `$\fun$' indicates a total function, which in this case models
an array. 

\paragraph{Coupling invariant.}
Now that we have defined the concrete type, we give a coupling invariant
that relates
a partial function $F$ to a hash table $H$:
\begin{equation}
\label{pfunhash:CI}
	H = makehash(F)
\end{equation}
where
\(
	makehash(F) = \{i: 0..N-1 @ (i, \Null)\} \oplus \{(K,V):F @ (hash(K), V)\}
\).
We have written
$makehash$ as
a \emph{set comprehension}.
In general, a set comprehension 
$\{x:T @ e(x)\}$ represents the set of values
of the expression $e(x)$ for each element $x$ of type $T$.
For example, $\{i: 0..N-1 @ (i, \Null)\}$ is the set of pairs 
$(i, \Null)$ for each number $i$ in the range $0..N-1$.
Thus $makehash(F)$ is a mapping from $hash(K)$ to $V$
for all pairs $(K,V)$ appearing in the function $F$, with all other
numbers mapping to $\Null$.
We assume we have available a module that implements operations
such as updates and accesses on arrays in constant time,
\eg, the \T{array} module in Mercury \cite{Mercury:95}.
We note the following property:
\begin{equation}
	F \in \PFun \land H = makehash(F) \imp H \in \Hash \label{imp:hash}
\end{equation}

\paragraph{\refcond{combined} for procedure update.}
As an example instantiation of \refcond{combined}, we prove that,
given the coupling invariant \refeqn{pfunhash:CI}, the
following procedure 
is a valid array implementation of $update$ (from \reffig{module:pfun}).
\[
    update \sdef (K, V, H, H') \prm \\
        \t1 \Ass{H \in \Hash \land K \in \sigma \land V \in \tau}, 
        \Spec{H' = H \oplus \{(hash(K), V)\}} 
\]
This can be implemented efficiently in Mercury by using the \T{set}
predicate from the \T{array} module.

First we show \refeqn{cond1:premise} entails \refeqn{proofob:noabort:p+}.
\[
    \CIHF \land F \in \PFun \land K \in \sigma \land V \in \tau \entails \\
        \t1 H \in \Hash \land K \in \sigma \land V \in \tau
\]
The conditions on $K$ and $V$ hold trivially, and $H \in \Hash$ follows
from \refeqn{imp:hash}.
We would normally expect \refeqn{proofob:noabort:p+} to be shown this easily.

Now we show the rest of \refcond{combined} holds.
\[
    \CIHF \land F \in \PFun \land K \in \sigma \land V \in \tau \entails \\
        \t1 (F' = F \oplus \{(K,V)\} \imp  \\
			\t2 (\exists H' @ H' = H \oplus \{(hash(K),V)\} \land \CIHFp)) \land \\
        \t1 (H' = H \oplus \{(hash(K),V)\} \imp  \\
			\t2 (\exists F' @ F' = F \oplus \{(K,V)\} \land \CIHFp))
\]
Simplifying using the one-point rule we get:
\[
    \CIHF \land F \in \PFun \land K \in \sigma \land V \in \tau \entails \\
        \t1 (F' = F \oplus \{(K,V)\} \imp  
			H \oplus \{(hash(K),V)\} = makehash(F')) \land \\
        \t1 (H' = H \oplus \{(hash(K),V)\} \imp  
			H' = makehash(F \oplus \{(K,V)\})) 
\]

Now we simplify the implications, combining them into
a single stronger predicate.
\[
    \CIHF \land F \in \PFun \land K \in \sigma \land V \in \tau \entails \\
        \t1 H \oplus \{(hash(K),V)\} = makehash(F \oplus \{(K,V)\})
\]
Thus we must show that, given that 
the coupling invariant for the inputs $F$ and
$H$ holds and that $F$ and the regular parameters are of the correct type, 
the coupling invariant holds for the output values.
We prove the conclusion by manipulating
the expression $H \oplus \{(hash(K),V)\}$.
\label{update:manipulation}
\begin{derivation}
	\step{H \oplus \HKV}
	\trans{=}{from antecedent $H = makehash(F)$; definition of $makehash$}
	\step{\{i: 0..N-1 @ (i, \Null)\} \oplus \{(X,Y):F @ (hash(X), Y)\} \oplus
		\HKV}
	\trans{=}{Since $F$ is a function and $hash$ is an injection}
	\step{\{i: 0..N-1 @ (i, \Null)\} \oplus 
		\{(X,Y):F\oplus \SKV @ (hash(X), Y)\}}
	\trans{=}{Definition}
	\step{makehash(F \oplus \{(K,V)\})}
\end{derivation}

Thus we have proved \refcond{combined} for $update$.
The full array implementation of the $\PFunName$ module is
shown later (\reffig{module:hash});
the remaining procedures are derived using
techniques described in \refsect{modderiv}.

\COMMENT{
\begin{derivation}
	\step{\{i: 0..N-1 @ (i, \Null)\} \oplus \{(X,Y):F \oplus \{(K,V)\} @ (hash(X), Y)\}}
	\trans{=}{since $F$ is a function, we extract the pair $(K,V)$}
	\step{\{i: 0..N-1 @ (i, \Null)\} \oplus \{(X,Y):F @ (hash(X), Y)\} 
		\oplus \{(X,Y):\SKV @ (hash(X), Y)\}}
	\trans{=}{simplify set comprehension for singleton set}
	\step{\{i: 0..N-1 @ (i, \Null)\} \oplus \{(X,Y):F @ (hash(X), Y)\} 
		\oplus \{(hash(K), V)\}}
	\trans{=}{from the assumption $CI(F,H)$}
	\step{H \oplus \{(hash(K), V)\}}
\end{derivation}
This is equal to the left-hand side of the equality, and the proof of
\refcond{combined} for $update$ is complete.
}

\COMMENT{
\subsection{Related work}
\label{mod:relwork}

Specifications of procedures and modules in our wide-spectrum
language are similar to Morgan's model-based module
specifications \cite{Morgan:94}, though in his case the modules provide a
`hidden' state that is not possible in traditional logic programs.
Bancroft and Hayes
\cite{BancroftHayes:RaMwOT}
have extended the imperative
calculus to include module specifications with opaque types similar to ours.
Our terminology is based on that of
Liskov and Guttag \cite{Liskov:86}, where a module
presents an interface of operations that can be performed on an opaque
data type.
Our module specifications are similar to the module declarations of
languages such as Mercury \cite{Mercury:95}.

There are many other existing logic programming frameworks for
modules or module-like encapsulation, \eg,
\cite{Specware:95,Ornaghi:96,Lau:99}.
Many of these define modules
by the algebraic specification
of abstract data types (ADTs)
\cite{Turski:87}.
The behaviour of a module is defined in
terms of the behaviour of its operations.  An implementation may be derived by
ensuring it maintains the behaviour of the module.
Read and Kazmierczak \cite{Read:92} present a particular method of
developing modular Prolog programs
from axiomatic specifications.
They write their programs in a
module system based on that of extended ML.
The specification of a module is written in the form of a set of
axioms stating the required properties of the procedures of the
module.
To define the semantics of
refinement, Prolog programs are considered to be equivalent to their
predicate completions.
The definition of module refinement in their approach is more general
than the technique presented in this paper:
any implementation that satisfies the axioms is valid
(\cf, interpretations between theories from logic \cite{Turski:87}).
However, for modules with a larger number of procedures,
presenting an axiomatic specification of how the procedures
interrelate is more problematic than with the model-based approach
used in this paper.
This is because axioms are required to define the possible interactions between
procedures, whereas, in the approach used in this paper, each procedure is
defined directly in terms of the model of the opaque type parameters.
In the algebraic approach, the proof of correctness amounts to showing that all
the axioms of the specification hold for the implementation \cite{Read:92}.
For a module with a large number of procedures this can be quite complex.
In comparison, the approach presented here
breaks down the problem into data refinement of each procedure
in isolation.
Although \refcond{combined} is quite complex, in practice, many of the
components of the condition are trivial 
and could be proved automatically.  
Additionally, in algebraic schemes, it is not clear how situations such as
that outlined in \refsect{equality-discussion} are handled;
either it is accepted that such programs may fail, or the allowed coupling
invariants are restricted to those with a one-to-one correspondence between 
abstract and concrete values.

Imperative data refinement \cite{Morgan:94} has more similarities with
our approach.  In that framework, a specification is augmented with the
concrete variable and the coupling invariant, then refinement proceeds
as normal, until the abstract variable is removed via diminution.
Neither of the augment and diminish steps are actual refinements, but
as in our framework the resulting relationship between the abstract and
concrete procedures is guaranteed to satisfy the conditions for $data$
$refinement$.

}

\COMMENT{
As shown by the set to list example,
the refinement relation, even with the
coupling invariant in context, is not enough to allow the efficient
implementations we can develop for programs in \opqf.
We have used the structure of the target program to allow us to make some
strong assumptions about the context.  
In fact, 
The sequential structure
we have chosen generalises the structure of imperative programming
languages, 
which implicitly enforce
the restrictions on how the procedures are to be called.
}

%% file: deriv.tex
\section{Calculating a concrete module}
\label{modderiv}

In the previous section we described the conditions that must hold between two
modules with respect to a coupling invariant to allow module refinement.
In this section we show how those conditions may be used
to calculate a concrete module,
given an abstract module and an appropriate coupling invariant.
The procedures of the calculated module are guaranteed to satisfy
\refcond{combined} with respect to their corresponding abstract procedures.
After introducing the general form of a calculated concrete procedure, we
specialise the technique 
based on the determinism of the coupling invariant and the abstract procedure.

\COMMENT{
The method we present will not always produce a concrete procedure,
since some coupling invariants are too weak to ensure the existence
of a concrete implementation.
It is also possible that the concrete procedure produced is not the
desired one, though in this case it is possible to rescue the
procedure without having to prove the full predicate of
\refcond{combined}.
}

\subsection{General form of concrete procedure}

The following theorem
gives the general form for the concrete procedure given the
abstract procedure and the coupling invariant.

\begin{Theorem}[Module calculation] 
\label{mod:deriv} 
Given a procedure $p \sdef (V,I,O) \prm \Ass{A},\Spec{P}$ 
with at most $V$ and $I$ free in $A$, and at most $V$, $I$ and $O$ free in $P$, 
if $p$ and the coupling invariant $CI$
satisfy 
the following properties for some predicate $R$ which is independent of $I$,
\begin{align}
	~& CI(I,I^+) \land A \land P \entails 
		(\exists O^+ @ CI(O,O^+)) \label{ci:check} \\
	~& CI(I,I^+) \land A \entails  \notag \\
		~& \t1 (\exists O @ P \land CI(O,O^+)) \iff R(V, I^+, O^+)
		\label{free:constraint}
\end{align}
then the following implementation of the concrete procedure
satisfies \refcond{combined}.
\begin{equation}
\begin{split}
\label{concrete:impl}
	~& p^+ \sdef (V, I^+, O^+) \prm  \\
		~& \t1\Ass{(\exists I @ CI(I,I^+) \land A)}, \\
		~& \t1 \Spec{(\all I @ CI(I,I^+) \land A \imp 
			(\exists O @ P \land CI(O,O^+)))}
\end{split}
\end{equation}
\end{Theorem}

With this theorem we may immediately derive a concrete module from an
abstract module that will satisfy \refcond{combined}, 
provided the coupling invariant satisfies \refeqn{ci:check} and
\refeqn{free:constraint}.

\Proof\footnote{
This is a simplified version of the proof that
originally appeared in \citeN{Thesis}.
}
To prove that \refeqn{concrete:impl} satisfies \refcond{combined},
we prove that it satisfies
\refeqn{proofob:noabort:p+}, \refeqn{proofob:step1} and
\refeqn{proofob:step5},
assuming
\refeqn{ci:check}, \refeqn{free:constraint},
\refeqn{cond1:premise}, and that the outputs $O$ and $O^+$ do not occur free
in the assumptions $A$ and $A^+$, respectively.

\COMMENT{
We also make use of the following predicate rule.
{\ReftLaw ~ \thlabel{imp:exists}}
\[
    \replace{E}{X}{T} \entails (\exists X @ E)
\]
This rules allows us to deduce that there exists some $X$ such that $E$ holds
if $E$ holds for some term $T$.
}

\begin{itemize}

\COMMENT{
\item[\refeqn{proofob:freeOutputs}]
The value for $A^+$,
given that $O \nfi A$, satisfies
\refeqn{proofob:freeOutputs}.
}

\item[\refeqn{proofob:noabort:p+}]

Substitute $(\exists I @ CI(I,I^+) \land A)$ for
$A^+$ in \refeqn{proofob:noabort:p+}; then
\refeqn{proofob:noabort:p+} holds from \refeqn{cond1:premise}.

\item[\refeqn{proofob:step5}]
Substituting
$(\all I @ CI(I,I^+) \land A \imp (\exists O @ P \land CI(O,O^+)))$
for 
$P^+$
in \refeqn{proofob:step5}, with
bound variable $I$ renamed to $X$, gives the following.
\[
    (\all X @ CI(X,I^+) \land \replace{A}{I}{X} \imp 
            (\exists O @ \replace{P}{I}{X} \land CI(O,O^+))) \imp \\
        \t1 (\exists O @ P \land CI(O,O^+))
\]
Since we have
$CI(I,I^+) \land A$
in context \refeqn{cond1:premise},
from the implication of the universally quantified predicate we may deduce
$(\exists O @ P \land CI(O,O^+))$.

\item[\refeqn{proofob:step1}]
Substituting
$(\all I @ CI(I,I^+) \land A \imp 
    (\exists O @ P \land CI(O,O^+)))$
for $P^+$
in \refeqn{proofob:step1}, with
variable renaming to avoid clashes, gives the following.
\[
    P \imp (\exists O^+ @ \\
        \t1 (\all X @ CI(X,I^+) \land \replace{A}{I}{X} \imp 
            (\exists Y @ \replace{P}{I,O}{X,Y} \land CI(Y,O^+))) \land \\
        \t1  CI(O,O^+))
\]

We simplify the middle line to $true$, assuming $P$ and $CI(O, O^+)$.
\begin{derivation}
	\step{\textstyle (\all X @ CI(X,I^+) \land \replace{A}{I}{X} \imp 
	            (\exists Y @ \replace{P}{I,O}{X,Y} \land CI(Y,O^+)))}
	\trans{\iff}{We use \refeqn{free:constraint} on 
		the quantification over $Y$.}
	\step{\textstyle (\all X @ CI(X,I^+) \land \replace{A}{I}{X} \imp 
	            R(V, I^+, O^+))}
	\trans{\iff}{Reduce the scope of $X$}
	\step{\textstyle (\exists X @ CI(X,I^+) \land \replace{A}{I}{X}) \imp 
	            R(V, I^+, O^+)}
	\trans{\iff}{$X$ is witnessed by $I$ from
		\refeqn{cond1:premise}}
	\step{\textstyle R(V, I^+, O^+)}
	\trans{\iff}{We now use \refeqn{free:constraint} again, from
		\refeqn{cond1:premise} in context}
	\step{\textstyle (\exists Y @ \replace{P}{O}{Y} \land CI(Y,O^+))}
	\trans{\iff}{$Y$ is witnessed by $O$  from
		assumptions $P$ and $CI(O, O^+)$}
	\step{true}
\end{derivation}

With the middle line simplified, we are left with
\[
	P \imp (\exists O^+ @ CI(O, O^+))
\]
Using \refeqn{cond1:premise} from the context, this follows from
\refeqn{ci:check}.

\COMMENT{
\[
    P \imp (\exists O^+ @ \\
        \t1 (\all X @ CI(X,I^+) \land \replace{A}{I}{X} \imp 
        R(V, I^+, O^+)) \land \\
        \t1  CI(O,O^+))
\]
We reduce the scope of $X$ because it does not occur free in $R(V, I^+, O^+)$.
\[
    P \imp (\exists O^+ @ \\
        \t1 ((\exists X @ CI(X,I^+) \land \replace{A}{I}{X}) \imp 
        R(V, I^+, O^+)) \land \\
        \t1  CI(O,O^+))
\]
Now we may eliminate the existential quantification over $X$ (choosing $X$
to be $I$) using \refeqn{cond1:premise}.
\[
    P \imp (\exists O^+ @ \\
        \t1 R(V, I^+, O^+) \land \\
        \t1  CI(O,O^+))
\]

We now use \refeqn{free:constraint} on $R$ again, from
\refeqn{cond1:premise} in context.
\[
    P \imp (\exists O^+ @ \\
        \t1 (\exists Y @ \replace{P}{O}{Y} \land CI(Y,O^+)) \land \\
        \t1  CI(O,O^+))
\]
Now $P$ and $CI(O,O^+)$ are in the context of the
quantification over $Y$, therefore the truth of the
existential quantifier is witnessed by $O$.
\[
    P \imp (\exists O^+ @ CI(O,O^+))
\]
}

\end{itemize}
\proofbox

The first assumption \refeqn{ci:check} of 
\refth{mod:deriv}
requires that the effect of the
abstract procedure implies that its output, $O$, has some concrete
representation.
This is typically just a type check on $O$, since it is the only free variable
in $(\exists O^+ @ CI(O, O^+))$.
One would always expect \refeqn{ci:check} to hold,
and in general it can be trivially discharged.
The second assumption \refeqn{free:constraint}
requires that the expression $(\exists O @ P \land
CI(O,O^+))$, in a context including $CI(I,I^+) \land A$, has
some equivalent form $R$ that does not include a free occurrence of
the abstract input $I$. 
In practice, one does not need to explicitly discharge
\refeqn{free:constraint}.
The given form of the concrete procedure \refeqn{concrete:impl} still
involves the abstract type via the coupling invariant (on both the
input and output).
One will need to simplify the concrete procedure to remove the
abstract data type;
once the abstract input has been removed (if possible),
\refeqn{free:constraint} has been satisfied.

Both constraints \refeqn{ci:check} and \refeqn{free:constraint} 
can be used as a consistency check for the
entire abstract module and chosen coupling invariant, prior to 
calculating the concrete module.
As mentioned, condition \refeqn{ci:check} fails for a
procedure $p$ if $p$ does not maintain the abstract type for its
output as expected by the coupling invariant.
Condition \refeqn{free:constraint} 
fails when information in the abstract type is lost in the
transformation to the concrete type, and the abstract procedures make use
of that information.
For instance, consider refining an ``abstract'' list module to a
``concrete'' set module, where the coupling invariant is just that the set
holds all the elements in the list (thus losing information about how many
times an element appears in the list, and the order of elements in the
list).
We can implement procedures for adding elements and checking membership
easily, however we would not expect to be able to implement a $count$
procedure, which returns the number of times an element appears in the list.  
Accordingly, we will not be able to prove \refeqn{free:constraint} for the
$count$ procedure with the chosen coupling invariant and concrete type.

%

In general, a carefully chosen coupling
invariant will ensure \refeqn{ci:check} and \refeqn{free:constraint}
hold.
In particular,
a coupling invariant in which
the abstract value is some function of the
concrete, \ie,  
$I = af(I^+)$,
will always ensure that \refeqn{free:constraint} holds.
This is because all occurrences of the abstract input can be replaced with
$af(I^+)$.


\subsubsection{Simplifying the specification}
\label{simp:spec}

In practice, the calculated specification of a concrete procedure
will be simpler than the general form given in \refeqn{concrete:impl}.
From \refeqn{free:constraint} we know that the right-hand side of the
implication can be expressed as $R(V,I^+,O^+)$, which does not contain $I$ or
$O$ free.
Making this simplification, and reducing the scope of $I$ gives
the following simpler specification part for $p^+$ in
\refeqn{concrete:impl}.
\[
	\Spec{(\exists I @ CI(I,I^+) \land A) \imp R(V,I^+,O^+)}
\]
The left-hand side of the implication matches the assumption from
\refeqn{concrete:impl}.  
Using the assumption and
\reflaw{equiv:specs:wrt} the specification may be simplified to just
\[
	\Spec{R(V,I^+,O^+)}
\]
Thus, in practice, once the specification has been calculated, it is just a
matter of simplifying $(\exists O @ P \land CI(O,O^+))$ to eliminate
references to $I$.
Then the universal quantification over $I$ becomes redundant.

\subsubsection{Initialisation and observer}

The following are instances of \refeqn{concrete:impl} simplified for
initialisations (no opaque inputs) and observers (no opaque outputs), 
respectively.
\begin{align}
	~& \Ass{A}, 
	\Spec{(\exists O @ P \land CI(O,O^+))} \label{impl:init} \\
	~& \Ass{(\exists I @ CI(I,I^+) \land A)}, 
	\Spec{(\all I @ CI(I,I^+) \land A \imp P)} \label{impl:obs}
\end{align}
\COMMENT{
\begin{align}
	~& \Ass{A}, 
	\Spec{(\exists O @ P \land CI(O,O^+))} \label{impl:init} \\
	~& \Ass{(\exists I @ CI(I,I^+) \land A)}, \notag \\
	~& \Spec{(\all I @ CI(I,I^+) \land A \imp P)} \label{impl:obs}
\end{align}
}
A consequence of there being no inputs for initialisations is that
\refeqn{free:constraint} can be trivially satisfied 
by choosing $R$ to be $(\exists O @ P \land CI(O,O^+))$.
Since there are no outputs for an observer
there is no need to check \refeqn{ci:check}.

\subsection{Example}
\label{update:deriv}

In \refsect{mr-example} we provided a proof 
that a concrete implementation of $update$ from \reffig{module:pfun}
satisfies \refcond{combined}.
Here we use \refth{mod:deriv} 
to calculate an implementation from the abstract $update$ procedure
and coupling invariant \refeqn{pfunhash:CI}.
We assume that \refeqn{ci:check} and \refeqn{free:constraint} hold (an example
of discharging these formally will be shown later in
\refsect{side-conditions}).
The concrete procedure in the form of \refeqn{concrete:impl} is thus:
\[
	\Ass{(\exists F @ \CIHF \land 
		F \in \PFun \land K \in \sigma \land V \in \tau)}, \\
	\Spec{(\all F @ \CIHF \land 
		F \in \PFun \land K \in \sigma \land V \in \tau \imp \\
			\t1 (\exists F' @ \CIHFp \land F' = F \oplus \SKV)) }
\]

From \refeqn{imp:hash} we may use \reflaw{weaken:ass} to refine
the calculated assumption.
\[
	\Ass{H \in \Hash \land K \in \sigma \land V \in \tau}
\]

We simplify the specification by applying the one-point law to $F'$.
\[
	\Spec{(\all F @ \CIHF \land 
		F \in \PFun \land K \in \sigma \land V \in \tau \imp \\
			\t1 H' = makehash(F \oplus \SKV)) }
\]
In \refsect{update:manipulation} we showed
\[
	makehash(F \oplus \SKV) = H \oplus \HKV
\]
Therefore we may rewrite the bottom line of the specification as
\[
	H' = H \oplus \HKV
\]
We have eliminated references to the abstract input
$F$ on the right-hand side of the
implication.
We therefore employ the simplification mentioned in \refsect{simp:spec}
to eliminate the quantification over $F$, resulting in the following
program
(identical to that given in \refsect{mr-example}).
\[
	\Ass{H \in \Hash \land K \in \sigma \land V \in \tau}, 
	\Spec{H' = H \oplus \HKV}
\]

\subsection{Specialisations}
\label{sect-specialisations}

In this section we provide some specialisations of \refeqn{concrete:impl},
based on the form of the coupling invariant and the abstract procedure.
The specialisations are partitioned based on two factors.
Firstly,
whether or not the abstract procedure $\Ass{A}, \Spec{P}$
is deterministic.
In a deterministic procedure there is only one possible
abstract output value given any 
regular and input parameter values, \ie, $P$ is of the form $O = f(V,I)$.
In a nondeterministic procedure, there could be many possible
output values related to any given regular regular and input values
(which we therefore write as a ternary relation $P(V,I,O)$).

Secondly, we partition the specialisations
based on the form of the coupling invariant.
We consider the case where the abstract variable is some (abstraction)
function of the concrete, $I = af(I^+)$.  In this situation, there are
potentially many concrete representations of an abstract value, though each
concrete value represents exactly one abstract value.
This is a common form of coupling invariant, and often simplifies the data
refinement process.
The second form of coupling invariant we consider is when the concrete
variable is some (concretisation) function of the abstract variable,
$I^+ = cf(I)$.
Thus each concrete value may represent many abstract values, though each
abstract value has exactly one concrete representation.
Finally we consider the case where the coupling invariant is a relation
between the abstract and concrete variables, $CI(I, I^+)$.

The specialisations are summarised in \reffig{specialisations}.
The predicates in the cells of the table are obtained from
\refeqn{concrete:impl} by simplifying using the one-point rule.
Most of the transformations are straightforward, however the case where the
abstract procedure is deterministic and the coupling invariant involves an
abstraction function is discussed in more detail in \refsect{demonic-deriv}.

\newcommand \multibox[2]{\begin{array}{l}#1\\\t1 #2\end{array}}

\begin{figure}[ht]
\begin{center}
\begin{tabular}{|l|c|c|}
\multicolumn{1}{c}{}
& \multicolumn{2}{c}{\emph{Abstract procedure}}
\\ \cline{1-3} 
\emph{Coupling}
& Deterministic & Non-deterministic \\
\emph{invariant}
& $O = f(V,I)$ & $P(V,I,O)$ \\
\cline{1-3} 
$I = af(I^+)$ 
	& $af(O^+) = f(V, af(I^+))$
	& $P(V, af(I^+), af(O^+))$ \\
$I^+ = cf(I)$ 
	& $\multibox{(\all I @ I^+ = cf(I) \land A \imp}{O^+ = cf(f(V,I)))}$ 
	& no simplification \\
$CI(I,I^+)$ 
	& $\multibox{(\all I @ CI(I,I^+) \land A \imp}{CI(f(V,I), O^+))}$
	& no simplification \\
\cline{1-3}
\end{tabular}
\caption{Specialisations for concrete specification $P^+$}
\label{specialisations}
\end{center}
\end{figure}

\subsection{Example: hash table}

In this section we use the derivation process to derive an array
($\Hash$) implementation of the abstract partial function type given
in \reffig{module:pfun}.
Recall the coupling invariant:
\[
	H = \{i: 0..N-1 @ (i, \Null)\} \oplus \{(K,V):F @ (hash(K), V)\}
	& \refeqn{pfunhash:CI}
\]
This coupling invariant is a concretisation function (the concrete variable
$H$ is a function of the abstract variable $F$).
There is an equivalent \emph{abstraction} function form (see
\refeqn{pfunhash:CI:2} below), 
but we prefer to use \refeqn{pfunhash:CI} for
the simplifications it provides in the calculation process.
When the abstract procedure is deterministic
we may use the simplification from the second row of \reffig{specialisations}.
The calculated procedures can all be implemented efficiently in the logic
programming language Mercury.

\subsubsection{Side conditions}
\label{side-conditions}
Before beginning the derivation, we check that the conditions 
\refeqn{ci:check} and \refeqn{free:constraint} hold for each procedure
in the module.
Condition \refeqn{ci:check} requires that the coupling invariant on the inputs,
as well as $A$ and $P$ of the abstract procedures, imply 
$(\exists O^+ @ CI(O,O^+))$.
Instantiating the quantification for the hash table example gives the
following condition which trivially holds:
\[
	(\exists H @ H = \{i: 0..N-1 @ (i, \Null)\} \oplus \{(K,V):F @
	(hash(K), V)\})
\]
It is easily seen that each abstract procedure guarantees that
the type of its output parameter is of type $\PFun$, and therefore
\refeqn{ci:check} holds for all the procedures in the module.

To satisfy \refeqn{free:constraint} we must be able to eliminate references
to the abstract type.
As mentioned earlier, this side condition is normally satisfied
implicitly in the derivation process, since in any case
we wish to eliminate the abstract variable.
However, we note that in this case
there is an equivalent coupling invariant that
we could employ:
\begin{equation}
\label{pfunhash:CI:2}
	F = (\lambda K: \sigma | H(hash(K)) \neq \Null @ H(hash(K)))
\end{equation}
This coupling invariant expresses the abstract variable $F$ as a
function of the concrete variable $H$.
The function is constructed by taking all keys $K$ of type $\sigma$ which are
not mapped to null by the hash table $H$ (the notation `$|$' is used
to restrict the domain of a function); all such keys are then mapped to
their (non-null) value in the hash table.
Because
the relationship between the abstract and concrete variables is one-to-one,
references to the abstract input can always be eliminated by
replacing them with the right-hand side of the equality in
\refeqn{pfunhash:CI:2}.
We may therefore automatically discharge \refeqn{free:constraint} for each
procedure in the module.

\subsubsection{Assumptions}
In the partial function module, the assumptions of the procedures are that the
input and regular parameters are of the correct type.
From \refeqn{concrete:impl} we calculate the concrete assumption
for the $update$ procedure.
\[
	(\exists F @ H = makehash(F) \land
		F \in \PFun \land K \in \sigma \land V \in \tau)
\]
From \refeqn{imp:hash} we may use \reflaw{weaken:ass} to refine
the calculated assumption.
\[
	H \in \Hash \land K \in \sigma \land V \in \tau
\]
Using similar manipulation, 
the assumption of each
concrete procedure is refined to 
the corresponding abstract assumption, except with 
$H \in \Hash$ in place $F \in \PFun$.
We now calculate  the specification of each concrete procedure, and refine
the specification to code (with the exception of
$update^+$ which was dealt with in \refsect{update:deriv}).

\subsubsection{Procedure init}
Since the $init$ procedure is a deterministic initialisation and the 
coupling invariant \refeqn{pfunhash:CI} is also deterministic,
we may immediately use the simplification in the second row of
\reffig{specialisations} with $f(V,I) = \es$.
Furthermore there are no inputs, eliminating the quantification over $I$.
\[
	H' = \{i: 0..N-1 @ (i, \Null)\} \oplus \{(K,V):\es @ (hash(K), V)\}
\]
The rightmost set comprehension is just the empty set, and therefore
the function override has no effect.
\[
	H' = \{i: 0..N-1 @ (i, \Null)\} 
\]
In other words, every element in the array is initialised to $\Null$.

\COMMENT{
\subsubsection{$update$}
We have already derived $update$ in \refsect{update:deriv}.
}

\subsubsection{Procedure remove}
This is a deterministic procedure, and we use the
specialisation in the second row of \reffig{specialisations}.
In this case $f(V,I)$ is $\{V\} \dsub I$.
\[
	(\all F @ \CIHF \land F \in \PFun \land K \in \sigma \imp \\
		\t1 H' = \{i: 0..N-1 @ (i, \Null)\} \oplus 
			\{(X,Y):\{K\} \dsub F @ (hash(X), Y)\})
\]
We simplify the equality on the bottom line.
Since $F$ is a function and $hash$ is an injection, it is equivalent to
\[
	H' = \{i: 0..N-1 @ (i, \Null)\} \oplus 
		(\{hash(K)\} \dsub \{(X,Y):F @ (hash(X), Y)\}
\]
Therefore $hash(K)$ must map to $\Null$ in $H'$.
\[
	H' = (\{i: 0..N-1 @ (i, \Null)\} \oplus{} \\
		\t1 \{(X,Y):F @ (hash(X), Y)\}) \oplus \{(hash(K), \Null)\}
\]
We rewrite using the $makehash$ function, and make the antecedent explicit
again.
\[
	(\all F @ \CIHF \land F \in \PFun \land K \in \sigma \imp \\
		\t1 H' = makehash(F) \oplus \{(hash(K), \Null)\})
\]
From the antecedent we replace $makehash(F)$ with $H$,
eliminating the reference to the abstract input $F$ on the right-hand side of
the implication.
We therefore use the simplification in \refsect{simp:spec} to eliminate the
quantification over $F$ and complete the refinement.
\[
	H' = H \oplus \{(hash(K), \Null)\}
\]

\subsubsection{Procedure access}

Since $access$ is an observer we instantiate \refeqn{impl:obs}.
\[
	(\all F @ \CIHF \land F \in \PFun \land K \in \sigma \imp \\
		\t1 K \in \dom(F) \land V = F(K))
\]
We manipulate the bottom line.
\[
	K \in \dom(F) \land V = F(K)
\]
Given the assumptions and $K \in \dom(F)$, $F(K) = H(hash(K))$.
\[
	 K \in \dom(F) \land V = H(hash(K))
\]
We have that $K \in \dom(F)$ is equivalent to $H(hash(K)) \neq \Null$
using \refeqn{pfunhash:CI:2}.
\[
	H(hash(K)) \neq \Null \land V = H(hash(K))
\]
We simplify.
\[
	V \neq \Null \land V = H(hash(K))
\]
As with $remove$, we have eliminated the abstract input from the conclusion of
the implication, and therefore eliminate the quantification over $F$ as in
\refsect{simp:spec}.
As expected, the procedure fails rather than
return $\Null$ for $V$ when $K$ is not in the domain.
The full module is given in \reffig{module:hash}.

\begin{figure}
\[
	\Module\ \Hashtable \\
	\mtype\ \Hash \sdef 0..N-1 \fun (\tau \union \{\Null\}) \\
	\also 
	\begin{array}{rl}
	init: & H: \Hash_o \\
	add: & K: \sigma, V: \tau, H: \Hash_i, H': \Hash_o \\
	access: & K: \sigma, H: \Hash_i, V: \tau \\
	remove: & K: \sigma, H: \Hash_i, H': \Hash_o \\
	\end{array}
	\also
	\begin{array}{rl}
	init \sdef & H \prm \Spec{H = \{i: 0..N-1 @ (i, \Null)\}} \\
	update \sdef & (K, V, H, H') \prm \begin{array}[t]{l}
		\Ass{H \in \Hash \land K \in \sigma \land V \in \tau}, \\
		\Spec{H' = H \oplus \{(hash(K), V)\}} \end{array}\\
	access \sdef & (K, H, V) \prm \begin{array}[t]{l}
		\Ass{H \in \Hash \land K \in \sigma}, \\
		\Spec{V \neq \Null \land V = H(hash(K))} \end{array}\\
	remove \sdef & (K, H, H') \prm \begin{array}[t]{l}
		\Ass{H \in \Hash \land K \in \sigma}, \\
		\Spec{H' = H \oplus \{(hash(K), \Null)\}} \end{array} \\
	\end{array} \\
	\also
	\End
\]
\begin{quote}
Assume the constants $hash$ and $N$ such that
$hash$ uniquely maps elements of type $\sigma$ to a natural number
in the range $0..N-1$.
\end{quote}

\caption{Concrete partial function module}
\label{module:hash}
\end{figure}

\COMMENT{
\subsection{Other examples}
\label{other:examples}

This would be nice to include but would involve another
example

\subsubsection{Infeasible derivations}

An abstract procedure and a given coupling invariant are not always
guaranteed to have a corresponding concrete implementation.
For instance, consider adding the following procedure to the module
$Calculator$, which removes the most recently added element from the
list.
\[
	chop(L, L') \sdef \Ass{L \in \listnat \land L \neq \el},
		\Spec{(\exists H @ L = [H|L'])}
\]

Intuitively, there is no possible concrete implementation for this
operation, since we lose information about the individual elements of
the list when we represent it using the $\calc$ type.
Accordingly, we cannot prove \refeqn{free:constraint}, which we
instantiate below.
\[
	\CITCL \land L \in \listnat \land L \neq \el \imp \\
		\t1 \underline{((\exists L' @ (\exists H @ L = [H|L']) \land \CITCLP)} \\
		\t1 \iff R(V,(T,C),(T',C'))
\]
We simplify the underlined predicate with the intent of removing references to
$L$, assuming the left-hand side of the
implication.
\[
	(\exists H @ (T',C') = (T-H,C-1) \land (\exists L' @ L = [H|L']))
\]
The abstract input $L$ still appears in the existentially quantified
predicate over $L'$.
Since we are assuming $L$ is non-empty, we can simplify the
quantification to $H = L(1)$, but we still have a reference to $L$
(\ie, there is no equivalent predicate $S$ for which $L \nfi S$).
Either the operation $chop$ must be dropped from the module, or the
coupling invariant strengthened to keep track of the individual
elements in the list.
}

\COMMENT{
\subsubsection{Multiple opaque outputs}

In this section we consider a procedure that has multiple opaque outputs
corresponding to a single opaque input.  
The approach presented in this section
generalises an earlier approach \cite{modreft:lopstr00}, which was unable to
deal with multiple opaque outputs.

Consider the subset procedure for the $Set$ module.
\[
	subset(S_1, S_2) \sdef \Ass{S_2 \in \setnat},
		\Spec{S_1 \subseteq S_2}
\]
We treat the subset $S_1$ as an output.  
There are potentially many subsets, $S_1$,  of the input set $S_2$.

We derive the following implementation by using \refeqn{concrete:impl} and
simplifying.
The conditions \refeqn{ci:check} and \refeqn{free:constraint} may be
trivially satisfied.
\[
	\Ass{L_2 \in \listnat}, \\
	\Spec{(\all S_2 @ S_2 = \ran L_2 \land S_2 \in \setnat \imp
		(\exists S_1 @ S_1 = \ran L_1 \land S_1 \subseteq S_2))}
\]
We apply the one-point law twice and simplify using
\reflaw{equiv:specs}.
\[
	\Ass{L_2 \in \listnat},
	\Spec{\ran L_1 \subseteq \ran L_2}
\]
This is the expected outcome, and 
may be
refined to the usual implementation of subset on lists (\cf, \cite{Lau:97}).

The technique presented in \cite{modreft:lopstr00} would have resulted
in the following proof obligation.
\[
	S_2 = \ran L_2 \imp \\
		\t1 \ran L_1 \subseteq \ran L_2 \land S_1 \subseteq S_2 \imp \\
			\t2 S_1 = \ran L_1
\]
This does not hold, since two subsets
($\ran L_1$ and $S_1$) of a set ($S_2$) are not necessarily equal.
}

%% file: demonic-deriv.tex
\section{Non-determinism in module derivations }
\label{demonic-deriv}

\COMMENT{
Logic programming languages like Prolog take the \emph{don't know}
interpretation of nondeterminism \cite{kowalski79},  
where a logic program is understood 
in terms of \emph{all} instantiations that satisfy it.
This is different from the \emph{don't care}, or \emph{demonic},
interpretation 
of nondeterminism taken in, for example, imperative program
refinement and in concurrent logic programming languages
\cite{Shapiro-89}, where
a single, arbitrary solution is chosen
though multiple solutions may be possible.
}

When dealing with refinement of opaque types (in which the
representation of the opaque type is not directly visible), there may be
multiple concrete representations of an opaque type variable which are
equivalent in terms of the abstract specification.  Thus if we choose
any one of those representations, the behaviour of the operations on
that representation will meet the requirements of the abstract
specification.  Hence for an opaque variable only one representation
from a set of equivalent representations needs to be chosen.  This
corresponds to \emph{don't care} or \emph{demonic} nondeterminism.  At
the same time, the abstract specification of an operation may involve
\emph{don't know} nondeterminism, where multiple answers, provided via
regular (non-opaque) variables, are possible.  Hence in order to handle
the information hiding aspects of opaque variables within the logic
programming context we need a framework that handles both \emph{don't
know} and \emph{don't care} nondeterminism; not just \emph{don't know}
(as in standard logic programming) and not just \emph{don't care} (as in
concurrent logic programming \cite{Shapiro-89}).

In this section
we apply the basic principles of demonic nondeterminism to module 
calculation.
We apply them to a particular combination of abstract procedure and coupling
invariant, for which the calculation method presented in \refsect{modderiv}
leads to a procedure that may produce many different answers for the
concrete output parameters, though we ``don't care'' which one is chosen.
This reduction in nondeterminism (in the choice of concrete value) will 
typically lead to a more efficient concrete
module.

\COMMENT{
In \reffig{specialisations} we gave the general form for a concrete procedure
when the abstract procedure is deterministic and the coupling invariant is an
abstraction function.
It is the weakest possible implementation, in that it is a procedure that
returns every possible concrete representation of the abstract output.
While this is a valid implementation, in some situations it may
be inefficient.
In this section we introduce the notion of \emph{demonic nondeterminism},
which allows a program to choose from a set of possible answers, rather than
returning them all.
We then apply the basic principles of demonic nondeterminism to derive a
concrete procedure that is still general, but also allows efficient
implementations to be developed.
}

\subsection{Deterministic abstract procedure and an abstraction function}
\label{det-abs-ci}

Consider the specialisation in the top-left entry in
\reffig{specialisations}, where we have a deterministic abstract procedure
and an abstraction function as the coupling invariant.
The calculated value of $P^+$ will be
\begin{equation}
\label{top-left-spec}
	af(O^+) = f(V,af(I^+)) 
\end{equation}
For example, this specialisation can occur when
representing a set $S$ as a list $L$.
Assume the existence of a module providing the opaque type $set$ and some
basic operations on sets, including a procedure, $add$, for adding an element to
a set (such a module can be found in \citeN{modreft:lopstr00}).
\[
	add \sdef (E,S,S') \prm \Spec{S' = \{E\} \union S}
\]
To represent the set as a list we choose the
coupling invariant to be the abstraction function
$S = \ran(L)$, where `$\ran$' returns the range or
set of
elements in a list.
Using \refeqn{top-left-spec} we calculate the corresponding
concrete procedure.
\[
	add^+ \sdef (E, L, L') \prm \Spec{\ran(L') = \{E\} \union \ran(L) }
\]
This procedure outputs a list $L'$ such that the elements of $L'$ are the elements
of $L$ plus $E$.
While this is valid, there are an infinite number of such lists because
an element is not precluded from appearing multiple times in $L'$.
Typically this will not be a practical implementation of the $add$ procedure
for lists.

Intuitively, however, since there is exactly one abstract output, there need
only be one concrete output. 
In other words, the calculated value for $P^+$ should be of the form
$O^+ = U$, for some term $U$.
In fact, any term $U$ with $V$ and $I^+$ free that satisfies the following
condition validates $O^+ = U$ as an implementation for $P^+$.
\begin{equation}
\label{guard-U}
	af(U) = f(V,af(I^+)) 
\end{equation}
This may be proved by substituting $O^+ = U$ for $P^+$ in \refcond{combined}
and simplifying (strengthening).

In the set-as-list example, we require some value $U$ for $L'$ such that
$\ran(U) = \{E\} \union \ran(L)$.
One obvious choice is 
$[E|L]$.
Clearly, $\ran([E|L]) = \{E\} \union \ran(L)$.
Thus, we are free to implement the concrete version of $add$ as 
$\Spec{L' = [E|L]}$, which is a stronger constraint on $L'$ than that
calculated by \refeqn{top-left-spec}.
To formalise the choice for $U$ we introduce demonic nondeterminism.

\subsection{Demonic nondeterminism}
\label{demonic:nondet}

\COMMENT{
In \citeN{nondet:01} a construct called demonic nondeterministic choice is
added to the wide-spectrum language.  Given a demonic
choice between multiple
answers, an implementation may choose any one of them.  This is in
contrast to our usual notion of implementation, where a program must return
all possible answers of its abstract specification.  With the demonic choice
operator and its associated semantics and refinement laws given in
\cite{nondet:01}, the wide-spectrum language can express
both \emph{don't know} and \emph{don't care} interpretations of nondeterminism within a
single program.
The \emph{don't know} interpretation is the default, and an explicit
nondeterministic choice operator ($\sqcap$) is required to express
\emph{don't care} nondeterminism.
}

In \citeN{nondet:01} a demonic choice operator ($\sqcap$) and its
associated semantics and refinement laws are added to the refinement
calculus.  This allows the wide-spectrum language to express the
\emph{don't care} interpretation of nondeterminism using $\sqcap$, as
well as the  the default \emph{don't know} interpretation of
nondeterminism, within a single program.
To understand the difference,
consider the program $S \sqcap T$.
It may be implemented by either of the programs
$S$ or $T$, as
embodied in the following refinement laws:
\[
	S \sqcap T \refsto S \\
	S \sqcap T \refsto T 
\]
Note the difference with program disjunction, where $S \lor T$ must be
implemented by returning the answers for both $S$ and $T$.
For example, consider the program
\[
	\Spec{X = 0} \sqcap \Spec{X = 1}
\]
This program is implemented by either the program $\Spec{X = 0}$ 
or the program $\Spec{X = 1}$.
In contrast,
the program
$\Spec{X = 0} \lor \Spec{X = 1}$ is not; it must return both answers for
$X$.

The identity of demonic choice is the program $\Magic$, that is,
$(\Magic \sqcap S) = (S \sqcap \Magic) = S$.
It is the (unimplementable) program that refines all other pro\-grams.

We may generalise the binary operator: given a program $S$ 
with free variable $X$, 
the demonic choice between the set of programs formed by 
instantiating $S$ with each possible value of $X$ is given by
\[
	\gendemon X @ S(X)
\]
This program is refined by $S(U)$, for all terms $U$.
We may limit the range of $X$ by introducing a \emph{guard}.
A guarded command $S \guard T$ is $\Magic$ if $S$ fails, but behaves
like $S,T$ otherwise.
To restrict the range of $X$ to just those terms that satisfy some predicate
$P$, we write
\[
	\gendemon X @ \Spec{P(X)} \guard S(X)
\]
For example, a program that picks exactly one arbitrary 
element from a set $A$ and sets
some variable $Y$ to have that value is:
\[
	\gendemon X @ \Spec{X \in A} \guard \Spec{Y = X}
\]
This is in contrast to the program $\Spec{Y \in A}$, which binds $Y$ to every
element of $A$.

A generalised demonic choice over $P(X) \guard S(X)$ 
is implemented by $S(U)$ for all terms $U$ that satisfy $P(U)$.
This is embodied in the following refinement law.
\ReftLaw[Eliminate generalised demonic choice ]{
\label{gendemon-eliminate}
$\t1
	\Rule{P(U)}
	{(\gendemon X @ \Spec{P(X)} \guard S(X)) \refsto S(U)}
$
}

We may refine a program $D$ to a generalised demonic choice if, for all terms
$X$ such that $P(X)$ holds, $D$ is refined by $S(X)$.
This is expressed by the following refinement law.
\ReftLaw[Introduce generalised demonic choice]{ 
\label{gendemon-intro}
$\t1
	\Rule{(\all X @ P(X) \imp (D \refsto S(X))}
	{D \refsto (\gendemon X @ \Spec{P(X)} \guard S(X))}
$
}

\subsection{Demonic choice in module calculation}

When there is only one abstract output value for a procedure, \ie, when it
is deterministic, we will typically want the corresponding
concrete procedure to also be deterministic. 
In other words, when $p$ is of the form
$\Ass{A}, \Spec{O = f(V, I)}$ for some assumption
$A$ and function $f$,
the corresponding $p^+$ should be of the form
$\Ass{A^+},\Spec{O^+ = U}$, where $A^+$ is the
calculated assumption and $U$ is some term involving $V$ and $I^+$.
However, when the coupling invariant allows many concrete representations
of an abstract value, \ie, when the coupling invariant 
is an abstraction function of the form $I^+ = af(I)$, 
the applicable derivation specialisation (top left in
\reffig{specialisations}) is not deterministic for $O^+$.

We solve this problem using demonic nondeterminism.
Recall from \refsect{det-abs-ci}
that $\Spec{O^+ = X}$ is a valid implementation of $p^+$
for all terms $X$ that satisfy \refeqn{guard-U}.
Expressing this formally:
\[
	(\all X @ af(X) = f(V,af(I^+)) \imp 
		(p^+(V,I^+,O^+) \refsto \Ass{A^+}, \Spec{O^+ = X}))
\]

From this, using \reflaw{gendemon-intro} we may deduce 
\begin{equation}
\label{demonic-impl}
	p^+(V,I^+,O^+) \refsto
	(\gendemon X @ \Spec{af(X) = f(V,af(I^+))} \guard 
		\Ass{A^+}, \Spec{O^+ = X})
\end{equation}

This specification
allows more flexibility in the final implementation of the concrete
procedure than the specification originally calculated (top left in
\reffig{specialisations}).
The implementor may choose any
term $U$ such that
$af(U) = f(V,af(I^+))$, and from \reflaw{gendemon-eliminate} the actual
implementation of $p^+$ becomes $O^+ = U$.
Without the reduction of nondeterminism, the implementor must retain each
concrete value that corresponds to the abstract output.

In the set-as-list example,
we would instantiate \refeqn{demonic-impl} to
calculate the list implementation of $add^+$.
\[
	\gendemon X @ \Spec{\ran(X) = \{E\} \union \ran(L)} \guard
		\Spec{L' = X}
\]
To refine this to code we choose some value for $X$ that satisfies the guard.
An obvious choice is $[E|L]$.
It may be easily seen that
$\ran([E|L]) = \{E\} \union \ran(L)$, which is the proof obligation for applying
\reflaw{gendemon-eliminate}.
We can therefore implement $add^+$ as
$(E, L, L') \prm \Spec{L' = [E|L]}$.

\COMMENT{
The benefit of using demonic choice in this manner is that the choice of 
concrete representation may be made within the concrete module.
The calculation process provides the general conditions that must be achieved.
Without the use of the demonic choice, the representation must be chosen when
the calculation is being performed.
}

%% file: relwork.tex
\section{Related work}
\label{relwork}

There is a large body of work on the
deductive synthesis of logic programs,
a survey of which appears in \citeN{SynthCompLogic:04}.
Deductive synthesis is a method for deriving a logic
program from a specification,
similar to the refinement calculus approach.  
A specification is manipulated using deduction rules (that
are proved correct within the proof framework), until an executable program is
reached.  The various approaches to deductive synthesis vary mainly in their
specification language; however, most use first-order logic since this can
express both specifications and logic programming code.  One of the most
developed schemes for deductive synthesis is that of
\citeN{Lau:97}.
They introduce a specification framework, which
underlies the synthesis steps, providing axioms and derived
relations.

The main difference between most deductive synthesis approaches and
logic program refinement is the inclusion of assumptions in the wide-spectrum
language.  These act as preconditions, providing a context for refinement
steps.  \citeN{Lau:97} have a $conditional$
specification, which includes an input relation for a procedure (\eg, types,
modes) with respect to which the synthesis of the procedure can take place.
The refinement calculus generalises this by allowing an assumption (input
relation) for any arbitrary program fragment.
A further difference is that in
deductive synthesis the deduction rules are 
derived with the SLD computation rule
in mind.
Thus issues such as clause-ordering are dealt with
during the synthesis process.
The refinement approach defers such issues to a separate translation phase,
where a particular implementation language (and computational model) are
chosen and the wide-spectrum program is translated into code for that
language.
A translation scheme for Mercury programs \cite{Mercury:95} is described in
\citeN{codegen:01}.

Despite these differences, much of the work on logic program development in
the synthesis world should be applicable in the refinement calculus.
The refinement calculus work has focused mainly on
the process of developing logic programs, while much of the synthesis work has
been developing strategies for deriving programs given particular forms of
specification.
We expect that such strategies can be formulated as sequences of refinement
rules.

\COMMENT{
Data refinement of logic programs 
presents some interesting situations that are
not apparent in traditional imperative data refinement
\cite{Morgan:93}.
In the imperative refinement calculus, data refinement is mostly
performed on state local to a module, which hides the details of a data
structure by providing an interface of procedures that use the structure.
Thus the representation can be changed without affecting any programs that
use the module, as long as the interface remains the same.  In traditional
logic programming there is no notion of local state, and the only way to share
information between procedures is to pass it as parameters.  This makes
the state 
variables `visible', as opposed to `hidden' in the imperative calculus.
Thus within a logic program that uses procedures whose parameters have been 
data-refined, 
these procedure calls must be consistently replaced with calls to
the corresponding concrete procedures.  
A data refinement of a procedure may therefore
impact on any program that uses this procedure.  Of course, data
refinement will sometimes be used on a variable that is internal to a
procedure, and will therefore not affect the interface; this has close
parallels in the imperative refinement calculus.  
}

The examples in \refsect{context} draw on work on
Prolog program transformations \cite{Sterling:94}, in
particular, transformations between the Prolog types $list$ and
$difference\ list$ \cite{marriott:88}.  
The relationship between the list and difference list implementations of
$reverse$ 
may also be defined with respect to higher-order program
synthesis, as shown by
\citeN{Seres:00}.

Specifications of procedures and modules in our wide-spectrum
language (\refsect{modreft}) are similar to Morgan's model-based module
specifications for imperative programs \cite{Morgan:94}, 
though in his case the modules provide a
`hidden' state (rather than type),
which is not possible in traditional logic programs.
\citeN{BancroftHayes:RaMwOT}
have extended the imperative
calculus to include module specifications with opaque types similar to ours.
Our module specifications are similar to the module declarations of
languages such as Mercury \cite{Mercury:95}.

There are many other existing logic programming frameworks for
modules or module-like encapsulation, \eg,
\cite{Specware:95,Ornaghi:96,Lau:99}.
Many of these define modules
through the algebraic specification
of abstract data types (ADTs)
\cite{Turski:87}.
An implementation module may be derived by
ensuring it maintains the axioms of the ADT.
\citeN{Read:92} present a particular method of
developing modular Prolog programs
from axiomatic specifications.
They write their programs in a
module system based on that of extended ML.
The specification of a module is written in the form of a set of
axioms stating the required properties of the procedures of the
module.
To define the semantics of
refinement, Prolog programs are considered to be equivalent to their
predicate completions.
The definition of module refinement in their approach is more general
than the technique presented in this paper:
any implementation that satisfies the axioms is valid
(\cf, interpretations between theories from logic \cite{Turski:87}).
However, for modules with a large number of procedures,
presenting an axiomatic specification of how the procedures
interrelate is more problematic than with the model-based approach
used in this paper.
This is because axioms are required to define the possible interactions between
procedures, whereas, in the approach used in this paper, each procedure is
defined directly in terms of the model of the opaque type.
In the algebraic approach, the proof of correctness amounts to showing that all
the axioms of the specification hold for the implementation \cite{Read:92}.
For a module with a large number of procedures this can be quite complex.
In comparison, the approach presented here
breaks down the problem into data refinement of each procedure
in isolation.

Imperative data refinement \cite{Morgan:94} has more similarities with
our approach to module refinement.  
In that framework, a specification is augmented with the
concrete variable and the coupling invariant, then refinement proceeds
as normal, until the abstract variable is removed via diminution.
Neither of the augment and diminish steps are actual refinements, but
as in our framework the resulting relationship between the abstract and
concrete procedures is guaranteed to satisfy the conditions for data
refinement.

The calculational method for deriving a concrete module from an abstract
module and a coupling invariant in \refsect{modderiv} is similar in 
style to that presented by \citeN{Morgan-Gardiner:90}.
The calculated concrete
procedures can appear quite complex in both methods
(\refeqn{concrete:impl} in this paper and
Lemma 3 in \cite{Morgan-Gardiner:90}).
However in the common situation in which the coupling invariant is an
abstraction function, that is, the abstract value is a function of the
concrete value, the one-point rules can be applied to simplify the
calculated procedures to term replacements on the abstract procedure.
These simplifications can occur in both settings.  In
either case, the bulk of the work
revolves around eliminating the existentially
quantified abstract state, and hence many data refinement techniques
should be applicable in both settings.
In the terminology of \citeN{Morgan-Gardiner:90}, our calculated
concrete procedure is \emph{valid}, that is, it is a module refinement
of the abstract.  However, it is not \emph{general} (unlike the
imperative calculated concrete procedure), because there are other
valid concrete procedures that are not (algorithmic) refinements of the
calculated procedure.
This necessitated the introduction of demonic nondeterminsm into the
calculation process in \refsect{demonic-deriv}.


%% file: concl.tex
\section{Conclusions}

\COMMENT{
This paper has extended the refinement calculus for logic programs in the
areas of contextual and module refinement.
In \refsect{wspl} the wide-spectrum language and refinement laws were
introduced.
\refsect{context} examined contextual refinement of logic
programs.  We introduced and simplified refinement laws for some of
the constructs of the language.
We illustrated how context is used by implementing a list reversal
program using difference lists.
We discussed module refinement in
\refsect{modreft}.  Modules are
groups of procedures that operate on a common data type.  By considering the
context provided by programs that use a module, efficient
implementations may be developed
by changing the module's data type.  
In \refsect{modderiv}
a technique for calculating
concrete modules from abstract modules was presented,
and illustrated by deriving an efficient implementation of a partial function
as a hash table.
We extended the calculation process to include demonic nondeterminism in
\refsect{demonic-deriv}.
This allowed concrete procedures to be calculated that contain a choice over
possible implementations.
}

This paper has described a cohesive framework for contextual refinement,
module refinement, and the calculation of concrete modules.  Contextual
information simplifies the refinement process by allowing individual refinement
steps and proof obligations to operate on the predicate level, with
minimal reference to the structure of the program.
Contextual information is collected via monotonicity laws, which not
only simplifies proofs ``by-hand'', but can also be made transparent to
the user when using a refinement tool \cite{Hemer:01}.  
The contextual laws presented in \refsect{context}
have been used to develop a solution to the \emph{N-queens} problem
\cite[Chapter 4]{Thesis}, and also in the development of a
term unification algorithm \cite{DLPfSUSR}.
In this paper we make use of contextual information in providing laws
for module refinement and calculation in a more convenient form.

Modules are an extension of the refinement calculus that allows data
abstraction and encapsulation.  In \refsect{modreft} we investigated an
implementation of the module specifying a partial function type in
\reffig{module:pfun}.
The partial function module has also been used in the development of
a term unification algorithm \cite{DLPfSUSR}.
The module calculation approach in \refsect{modderiv} can be used to
automatically derive a concrete module from an abstract module and
coupling invariant.  The calculated module is guaranteed to satisfy the
conditions for module refinement, thus automatically discharging the
proof obligations associated with module refinement.  
However, while the calculated module is a valid module refinement, there
are in general many valid module refinements, some of which may be more
efficient than the calculated version.
This can occur in the common situation where the abstract procedure is
deterministic and there are many possible concrete representations of an
abstract value.  
To overcome this problem, in \refsect{demonic-deriv}
we introduced a demonic, or \emph{don't care},
nondeterministic operator into the calculation process.
This approach can be used to eliminate unwanted nondeterminism
introduced by the coupling invariant.


\COMMENT{
The resulting theory for contextual and module refinement
is both an extension and generalisation of earlier
work in the area.
The module calculation theory introduced in this paper
is a powerful method for finding implementations of
abstract modules. 
This was made possible by making assumptions about the context in which
modules are used.
The approach is not always guaranteed to produce an efficient implementation.
However, as shown in \refsect{demonic-deriv}, in some circumstances this
problem can be avoided through the use of demonic nondeterminism.

In other work \cite{DLPfSUSR} we have used the partial function module
given in \refsect{modreft} in the development of a non-trivial program for
unifying terms.  
As with the development of a solution to the N-queens
problem \cite{reflp:01} which made use of high-level
refinement rules for introducing recursion, 
it is apparent that the logic program
refinement calculus technique is better suited to the refinement of such
programs than imperative refinement calculi (an example development of
an imperative solution to the N-queens problem is given in \cite{Back:98}).
This is because the final logic programs are naturally closer to the
specification than the corresponding imperative programs are.
It is also the case that recursion, which is typically
fundamental to any non-trivial logic program implementation, is easier to
introduce in the logic programming refinement calculus than to introduce
iteration in an imperative program development, in which guards, invariants,
and variants must be considered.  Logic programming languages are naturally
suited to expressing recursion and recursive data types.
}

\COMMENT{
Demonic nondeterminism, as discussed in \refsect{demonic-deriv}, is an
example of ``don't care'' nondeterminism, as distinct from the usual
``don't know'' nondeterminism inherent in logic programming \cite{kowalski79}.
The difference between the two is also relevant in the field of concurrent
logic programming \cite{Shapiro-89}, where ``don't care'' nondeterminism is 
used for process synchronisation on partially instantiated variables.  
An interesting avenue of future work is to use a concurrent logic 
programming language as the target implementation language, especially
if we allow opaque input and output variables to act as
synchronisation variables.
This would 
potentially allow a more cohesive treatment of ``don't care''
nondeterminism at the implementation level, as well as simplify the
module calculation process in the circumstances where ``don't care''
nondeterminism is introduced by the coupling invariant.
}



\COMMENT{
The primary aim of this paper was to introduce a notion of data refinement in
the logic programming refinement calculus.  Data refinement is an important
facet of formal development, providing a method
for transforming abstract types used in specifications to implementation data
types in the final program.
The paper initially showed how this may be done on a procedure-by-procedure
basis, but this was extended to modules that export an opaque data type.
The introduction of opaque types allows a modular design in our specification
language, while considering data refinement at the module level allowed some
refinements that were not possible when procedures were considered in
isolation.

A secondary aim of the paper was to develop the
notion of contextual refinement, of which the data refinements presented 
are specialisations.
Contextual refinement allows a subprogram to be refined with respect to the
context in which
it appears; effectively, the structure of the top-level program is
summarised as a predicate (the context), which may be used to discharge proof
obligations in the refinement of the subprogram.
Thus refinement laws are simplified, since the majority need not contain
complex program structures to represent context.  
This method of context management is suited to both ``by-hand'' refinements and
tool-supported refinements.
The implicit construction and maintenance of context can be managed
by the tool, and the number of laws that must be used is significantly reduced
if context is managed in this way (since assumptions in general do not have to
be directly manipulated).
Contextual laws for the refinement calculus for logic programs have been
developed in a refinement tool \cite{Hemer:01}.
}

\COMMENT{
The use of \emph{parallel} conjunction for context 
less straightforward, but is useful when,
for instance, a type of a parameter is not assumed by a procedure,
but instead must be established.
In this paper we restrict ourselves to examples where only sequential
conjunction provides required context, and thus do not discuss the context
provided by parallel conjunction.
For the details, the reader is referred to \cite{Thesis}.

Thus we are dealing only with the most abstract form of programs, \ie, those
made up only of an assumption and effect component.
In \cite{modreft:lopstr00} we show how this can be generalised
to deal with any form of the
abstract and concrete procedures.

Despite that the calculation technique does not necessarily
lead one directly to the desired
implementation,
the work associated with calculating a module refinement will typically
be less than that associated with choosing an implementation then checking it
against \refcond{combined}.
}

\COMMENT{
\subsection{Future work}
\label{concl:future:work}

An avenue for future work is to incorporate modules, module refinement,
and data refinement into the existing logic programming refinement tool
\cite{Hemer:01}.
This tool is already capable of handling context in the manner described here.
The incorporation of module specifications should be straightforward, resulting
in a simple click to generate the proof obligations associated with two
modules (many of which should be automatically discharged).  
The tool could be extended to derive concrete modules from abstract
modules, by applying the process outlined in 
\refsect{modderiv}.
}

\COMMENT{
The three cases of data refinement discussed
in \refsect{datareft},
involving
the setting up of the coupling invariant as context, augmenting and
diminishing,
and the specification of auxiliary procedures, are all specialisations of
already existing laws in the tool.

Recent developments in the refinement calculus for logic programs project
include a code generator and a semantics for demonic nondeterminism.  
The former could be extended to accept module specifications and to generate
module code in some language, \eg, Mercury.  The latter would be useful in the
module refinement work, where we may refine a nondeterministic
abstract procedure to a deterministic concrete procedure.  At the moment this
is done via contextual refinement with regards to existentially quantified
outputs. 
In cases where there are many possible concrete representations
for an abstract value, 
a demonic choice operator may simplify the data refinement.
}

{\bf Acknowledgments.}  The authors would like to thank David Hemer and
three anonymous referees for constructive comments on earlier versions of
the paper.
The work reported in this paper was supported by
Australian Research Council grant number A49937007: \emph{Refinement
Calculus
for Logic Programming}.